\documentclass[conference]{sty/IEEETran}
\usepackage[square,comma,numbers,sort&compress]{natbib}

\let\labelindent\relax

\usepackage{xcolor}
\usepackage{xurl}
\usepackage{hyperref}

\usepackage{amsmath,amsopn,amssymb}
\usepackage{subfigure}
\usepackage{endnotes,microtype,xspace,graphicx,fancyvrb,multirow}
\usepackage{booktabs}
\usepackage{array,underscore,relsize}
\usepackage[T1]{fontenc}
\usepackage{times}
\usepackage{fancyhdr}
\usepackage{enumitem}
\usepackage[labelfont=bf,font=small,skip=5pt]{caption}
\usepackage{pifont}
\usepackage{comment}
\usepackage{tikz, pgfplots}
\usepackage{makecell}
\usepackage{xstring}

\usepackage[linesnumbered,ruled,vlined]{algorithm2e}

\usepackage{fp}
\usepackage{siunitx}

\usepackage[frozencache,cachedir=minted-cache]{minted}

\usepackage{balance}

\usepackage[title]{appendix}

\newcommand{\sys}{\mbox{\textsc{Prime+Retouch}}\xspace}
\newcommand{\TSX}{\mbox{Intel TSX}\xspace}
\newcommand{\SGX}{\mbox{Intel SGX}\xspace}

\newcommand{\retouchonly}{\mbox{\textsc{Retouch}}\xspace}
\newcommand{\primeprobe}{\mbox{\textsc{Prime+Probe}}\xspace}
\newcommand{\flushreload}{\mbox{\textsc{Flush+Reload}}\xspace}

\newcommand{\primeabort}{\mbox{\textsc{Prime+Abort}}\xspace}
\newcommand{\reloadrefresh}{\mbox{\textsc{Reload+Refresh}}\xspace}

\definecolor{color1}{HTML}{FF496C}
\definecolor{color2}{HTML}{45CEA2}
\definecolor{color3}{HTML}{2B6CC4}

\definecolor{lcolor1}{HTML}{8dd3c7}
\definecolor{lcolor2}{HTML}{ffffb3}
\definecolor{lcolor3}{HTML}{bebada}
\definecolor{lcolor4}{HTML}{fb8072}
\definecolor{lcolor5}{HTML}{80b1d3}
\definecolor{lcolor6}{HTML}{fdb462}
\definecolor{lcolor7}{HTML}{b3de69}
\definecolor{lcolor8}{HTML}{fccde5}
\definecolor{lcolor9}{HTML}{d9d9d9}
\definecolor{lcolor10}{HTML}{bc80bd}
\definecolor{lcolor11}{HTML}{ccebc5}
\definecolor{lcolor12}{HTML}{ffed6f}
\definecolor{lcolorbot}{HTML}{95918c}

\begin{document}
\renewcommand{\ttdefault}{pxtt}

\let\URL\relax
\newcommand{\URL}{\url}
\newcommand{\cc}[1]{\mbox{\smaller[0.5]\texttt{#1}}}

% enable the below for ACM camera ready
%\clubpenalty=10000
%\widowpenalty=10000
%\renewcommand*{\bibfont}{\raggedright}

%\linespread{1.2}

\fvset{fontsize=\scriptsize,xleftmargin=8pt,numbers=left,numbersep=5pt}

\input{code/fmt}
\newcommand{\figrule}{\hrule width \hsize height .33pt}
\newcommand{\coderule}{\vspace{-0.5em}\figrule\vspace{0.2em}}

\setlength{\abovedisplayskip}{0pt}
\setlength{\abovedisplayshortskip}{0pt}
\setlength{\belowdisplayskip}{0pt}
\setlength{\belowdisplayshortskip}{0pt}
\setlength{\jot}{0pt}

% change the below value to decrease the gap between fig/tab and texts
% \setlength{\textfloatsep}{1em}
\setlength{\textfloatsep}{0.5em}

\def\Snospace~{\S{}}
\renewcommand*\sectionautorefname{\Snospace}
\def\sectionautorefname{\Snospace}
\def\subsectionautorefname{\Snospace}
\def\subsubsectionautorefname{\Snospace}
\def\chapterautorefname{\Snospace}
\newcommand{\subfigureautorefname}{\figureautorefname}

%\numberwithin{equation}{section}
\newcommand{\yes}{Y}
\newcommand{\no}{}

% sema
\newcommand{\shl}{\ \cc{<}\cc{<}\ }
\newcommand{\shr}{\ \cc{>}\cc{>}\ }
\newcommand{\x}{$\times$\xspace}

\if 0
\renewcommand{\topfraction}{0.9}
\renewcommand{\dbltopfraction}{0.9}
\renewcommand{\bottomfraction}{0.8}
\renewcommand{\textfraction}{0.05}
\renewcommand{\floatpagefraction}{0.9}
\renewcommand{\dblfloatpagefraction}{0.9}
\setcounter{topnumber}{10}
\setcounter{bottomnumber}{10}
\setcounter{totalnumber}{10}
\setcounter{dbltopnumber}{10}
\fi

\newif\ifdraft\drafttrue
\newif\ifnotes\notestrue
\ifdraft\else\notesfalse\fi

% hide comments
% \renewcommand{\TK}[1]{\ignorespaces}
% \renewcommand{\XXX}[1]{\ignorespaces}
% \renewcommand{\TODO}[1]{\ignorespaces}

%% Ensure ligatures (e.g., ``fine official flag'') can be copy/pasted from PDF.
\input{glyphtounicode}
\pdfgentounicode=1

\newcolumntype{R}[1]{>{\raggedleft\let\newline\\\arraybackslash\hspace{0pt}}p{#1}}

% include macros
\newcommand{\includepdf}[1]{
  \includegraphics[width=\columnwidth]{#1}
}
\newcommand{\includeplot}[1]{
  \resizebox{\columnwidth}{!}{\input{#1}}
}

% list
\newcommand{\squishlist}{
\begin{itemize}[noitemsep,nolistsep]
  \setlength{\itemsep}{-0pt}
}
\newcommand{\squishend}{
  \end{itemize}
}

%%
%% NOTE.
%%  to use circled number in caption, use
%%   (e.g., \protect\WC{1})
%%
\newcommand*\WC[1]{%
\begin{tikzpicture}[baseline=(C.base)]
\node[draw,circle,inner sep=0.2pt](C) {#1};
\end{tikzpicture}}

\newcommand*\BC[1]{%
\begin{tikzpicture}[baseline=(C.base)]
\node[draw,circle,fill=black,inner sep=0.2pt](C) {\textcolor{white}{#1}};
\end{tikzpicture}}

\newcommand{\PP}[1]{
\vspace{2px}
\noindent{\bf \IfEndWith{#1}{.}{#1}{#1.}}
}

\newcommand{\PN}[1]{
\vspace{2px}
\noindent{\bf #1}
}

\newcommand{\ra}[1]{\renewcommand{\arraystretch}{#1}}
\renewcommand{\O}{\phantom{0}}

%% units
\newcommand{\B}{\,\text{B}\xspace}
\newcommand{\M}{\,\text{M}\xspace}
\newcommand{\T}{\,\text{T}\xspace}
\newcommand{\KB}{\,\text{KB}\xspace}
\newcommand{\MB}{\,\text{MB}\xspace}
\newcommand{\GB}{\,\text{GB}\xspace}
\newcommand{\TB}{\,\text{TB}\xspace}

\newcommand{\Bs}{\,\text{B/s}\xspace}
\newcommand{\KBs}{\,\text{KB/s}\xspace}
\newcommand{\MBs}{\,\text{MB/s}\xspace}
\newcommand{\GBs}{\,\text{GB/s}\xspace}

% boxbeg/end
\newcommand{\boxbeg}{
\vspace{2px}
\noindent\begin{tabular}{|l|}\hline
\begin{minipage}{3.2in}
\vspace{2px}
\noindent
}

\newcommand{\boxend}{
\vspace{2px}
\end{minipage}\\ \hline
\end{tabular}
\vspace{-10pt}
}

% shortcuts
\newcommand{\etal}{\textit{et al}.\xspace}
\newcommand{\ie}{\textit{i}.\textit{e}.}
\newcommand{\eg}{\textit{e}.\textit{g}.}
\newcommand{\aka}{\textit{a}.\textit{k}.\textit{a}\xspace}

\renewcommand{\algorithmautorefname}{Algorithm}

\gdef\therev{cd8331c}
\gdef\thedate{2022-09-22 08:25:04 -0400}

% \title{\sys: a paper template}

% % when 'make draft'
% \ifdefined\DRAFT
%  \pagestyle{fancyplain}
%  \lhead{Rev.~\therev}
%  \rhead{\thedate}
%  \cfoot{\thepage\ of \pageref{LastPage}}
% \fi

% % blind
% % \author{Paper \#123}

% \author{
%  Isaac Newton$^\dagger$\; 
%  Albert Einstein$^\ddagger$\;
%  Marie Curie$^\ast$\;
%  James Clerk Maxwell$^\dagger$\;
% \\
%  Richard Feynman$^\dagger$\;
% \\\\
%  \emph{$^\dagger$  Trinity College}, \\
%  \emph{$^\ddagger$ University of Zurich}, 
%  \emph{$^\ast$ 	 University of Paris}
% }

\title{\sys: \\ When Cache is Locked and Leaked}
\author{\IEEEauthorblockN{Jaehyuk Lee}
\IEEEauthorblockA{Georgia Institute of Technology\\
jaehyuk@gatech.edu}
\and
\IEEEauthorblockN{Fan Sang}
\IEEEauthorblockA{Georgia Institute of Technology\\
fsang@gatech.edu}
\and
\IEEEauthorblockN{Taesoo Kim}
\IEEEauthorblockA{Georgia Institute of Technology\\
taesoo@gatech.edu}}

% \IEEEoverridecommandlockouts
% \makeatletter\def\@IEEEpubidpullup{6.5\baselineskip}\makeatother
% \IEEEpubid{\parbox{\columnwidth}{
%     Network and Distributed System Security (NDSS) Symposium 2023\\
%     28 February - 4 March 2023, San Diego, CA, USA\\
%     ISBN 1-891562-83-5\\
%     https://dx.doi.org/10.14722/ndss.2023.23xxx\\
%     www.ndss-symposium.org
% }
% \hspace{\columnsep}\makebox[\columnwidth]{}}

\maketitle

\sloppy

\begin{abstract}
% Side channels exposed by 
% shared microarchitectural resources have become the 
% Achilles' heel of modern computer systems,
% as they permeate architectural security boundaries.
%
Caches on the modern commodity CPUs
have become one of the major sources
of side-channel leakages
and been abused as a new attack vector.
To thwart the cache-based side-channel attacks,
two types of countermeasures have been proposed:
detection-based ones
that limit the amount of microarchitectural traces an attacker can leave,
and cache prefetching-and-locking techniques
that claim to prevent such leakage
by disallowing evictions on sensitive data.
In this paper, we present the \sys attack that
completely bypasses these defense schemes
by accurately inferring the cache activities
with the metadata of the cache replacement policy.
\sys has three noticeable properties:
1) it incurs no eviction on the victim's data,
allowing us to bypass the two known mitigation schemes,
2) it requires minimal synchronization of
only one memory access to
the attacker's pre-primed cache lines,
and 3) it leaks data via non-shared memory, yet
because underlying eviction metadata is shared.

We demonstrate \sys in two architectures: predominant Intel x86 and emerging Apple M1.
We elucidate how \sys can break the T-table implementation of AES
with robust cache side-channel mitigations
such as Cloak, under both normal and SGX-protected environments.
We also manifest feasibility of the \sys attack on 
the M1 platform imposing more restrictions
where the precise measurement tools such as 
core clock cycle timer and performance counters are 
inaccessible to the attacker. 
Furthermore, we first demystify 
undisclosed cache architecture and its eviction policy 
of L1 data cache on Apple M1 architecture.
We also devise a user-space noise-free cache monitoring tool
by repurposing Intel TSX. 
\end{abstract}

\section{Introduction}
\label{s:intro}

Cache side-channel attacks have recently
gained increasing attention
due to their broad impacts~\cite{elgamal-attack, cache-cloud1, cache-cloud2, flush+reload, gruss2015cache, cachezoom,sge}.
The majority of cache attacks rely on
the observable timing differences between
a cache hit and a miss caused by
the access latency of memory hierarchies.
By carefully manipulating
a target cache set,
the attacker can force timing differences
that lead to leakage of the victim's secret data.

As cache side-channel attacks have
continuously broken carefully designed systems,
various detection and mitigation techniques targeting them
have been proposed~\cite{attack-survey}.
Practical and widely experimented defense mechanisms are based on the assumption
that cache side-channel attacks pose observable side effects themselves as well.
That is,
detectable attacker efforts,
commonly forcing cache evictions,
are required to capture meaningful
victim memory access activities.
Therefore,
several mitigation proposals
aim to detect such traces imprinted on the cache
by directly monitoring abnormal
microarchitectural behaviors~\cite{time-defense, spydetector, realtime-hpc, cacheshield, cloudradar}.
Others seek to conceal the victim's access pattern
by preloading all necessary data into the cache before 
issuing sensitive memory operations~\cite{wait-minute, fine-grain-cross, smokebomb, sbox-not-secure,evict+time}.
Recently~\cite{cloak} demonstrated that
locking the preloaded data in the cache
using transactional memory
can further eliminate cache evictions of preloaded data.
%While detection-based countermeasures
%restrict the amount of microarchitectural traces
%an attacker can leave,
%cache preloading and locking techniques~\cite{cloak}
%are promising to completely prevent
%attacks that cause cache misses on sensitive code and data.
%
% However,
% security research is all about
% continuously challenging the status quo.

In this paper,
we introduce a new attack, \sys,
% \XXX{connect with the prev par by saying,
%   completely bypass the known mitigation
% for cache side-channel attacks.}
that can leak information
while maintaining the victim cache state and minimizing microarchitectural traces,
completely bypassing the known mitigation techniques.
The core idea of \sys
is to accurately infer the cache accesses
from the metadata of the cache replacement policy.
% Tree-based Pseudo Least-Recently Used policy (Tree-PLRU).
% \XXX{More specifically, trim below:}
More specifically,
after reverse engineering
the Tree-based Pseudo Least-Recently Used replacement policy (Tree-PLRU)
of the Intel and Apple M1 processors,
%%
%we thoroughly analyze
%the characteristics of the policy
%and discover that by precisely manipulating L1 cache entries,
we discover that 
the attacker can learn the memory access history of a co-located victim 
from the states of the Tree-PLRU's tree-shaped metadata
without causing evictions of sensitive data and code.

% \XXX{BREAK!}
\sys has three unique properties
  compared with the popular forms of cache-based attacks~\cite{prime-probe-dcache,
    flush+reload}.
% We therefore present the \sys attack,
% a pioneer of the metadata-based cache attack family
% that leaks sensitive information through
% metadata of the cache replacement policy
% instead of through the cache state directly.
%
% The \sys attack
% stands out from timing-based attacks in that
1) no eviction of sensitive data and code is required,
2) no shared memory with the victim is required, and
3) the memory access pattern is leaked through the metadata of the cache replacement policy.
The \sys attack proves that
it is possible to distinguish actual memory accesses
without accessing or evicting preloaded data,
which completely breaks the assumption
of prefetch-based cache side-channel mitigations.
The \sys attack is also challenging to detect,
as the victim cannot
identify the source of cache accesses while
\sys only requires single synchronized cache access
to one of the attacker-primed entries.
%XXX{instead of hardware modification we need some more meaningful sentences to emphasize the impact of the attack}
Total mitigation might even require hardware modifications regarding the L1 cache replacement policy.
To our best knowledge,
the \sys attack is the first 
cache side channel attack that 
abuses L1 replacement policy
to leak the victim's access pattern
on Intel x86 and Apple M1 architectures.

However,
it is non-trivial to reliably demonstrate \sys
in a modern CPU for three reasons:
First,
as \sys does not directly rely on timing information,
it is hard for the attacker to synchronize with
the victim's operations
and interfere at the correct moment.
Second,
analysis activities during attacks
waste processor cycles
and further exacerbate
the above challenge.
Lastly,
as Tree-PLRU is sensitive to
the order of cache accesses,
unsynchronized cache accesses from the attacker
render undesired states of the Tree-PLRU metadata
that introduce no meaningful information except noise.

We successfully overcome those challenges
by postponing memory access analysis after the active attack phase using the Tree-PLRU metadata
and the novel technique of PLRU Aware \retouchonly~\autoref{ss:post-sync},
which is the first of its kind comparing to
noisy realtime measurements adopted by traditional side-channel attacks.
We demonstrate realistic \sys attacks in action
(\eg, attacking AES T-Table)
and show that the attack successfully bypasses
all prefetch-based countermeasures.
We further extend the \sys attack to Apple's M1 platform.
We have reported the \sys attack to Intel
and got acknowledged about the attack method.

\PP{Summary} This paper makes the following contributions:
\begin{itemize}[noitemsep,topsep=0pt]
	\item We introduce a novel method that allows noise-free L1 cache monitoring
		using Intel TSX in user space.
	\item We first study the cache architecture of Apple's M1 processor using
		the undocumented performance counter 
		and expose the underlying Tree-PLRU policy of its L1 cache.
	\item We propose \sys,
		a metadata-based L1 cache side-channel attack
		that only requires a single synchronized memory access without evicting the victim's data,
		and show how \sys completely breaks the assumption of prefetch-based mitigations.
	\item We provide thorough analysis of the unique properties of the Tree-PLRU eviction policy
        and develop techniques to enable the attacker to synchronize with the victim
        and make the \sys attack practical.
	\item We demonstrate that the \sys attack
		successfully leaks victim secrets (\eg, access patterns in AES T-Table)
		not only under settings assumed by traditional side-channel attacks, but also SGX-protected environments
	\item We demonstrate that the \sys attack
		can be extended to the M1 platform where no fine-grained timer is available
		by purely leveraging the Tree-PLRU policy. 
\end{itemize}

\if
Although logical level isolations, 
such as those provided by
operating systems and hypervisors,
ensure secure multi-threading,
processor microarchitectural resources
are still shared among logically isolated processes
to maximize task multiplexing and
optimize performance.
Unfortunately,
those shared resources
expose opportunities for attackers
to intentionally create contention and
infer useful information about the victim process
co-located within the same sharing domain~\cite{branch1, branch2, dram1, dram2, cache1, cache2, flush+reload}.
%
% Those side effects are usually not 
% discovered and fixed
% prior to the public release of units.
%
The so-called side-channel attacks
thus have become the Achilles' heel
of modern computer systems,
as they are able to
permeate architectural security boundaries.
\fi

\if 0
On one hand,
adopting hardware countermeasures
is too restricted by manufacturers
and takes years to deploy~\cite{random-fill-cache, pl-rp-cache}.
%
On the other hand,
performance overhead caused by
operating system or hypervisor modifications~\cite{stealthmem, cachebar, catalyst}
is too expensive.
%
\fi

\section{Background}
\label{s:background}

\subsection{Cache Architecture}
\label{cache}
CPU caches are small but fast
memory chunks located between
CPU cores and RAM.
%
%As such memory chunks are placed
%directly on die and close to cores,
%they impose significantly lower access latency
%compared to RAM
%%
It serves to store data
that the CPU is most likely to need next.
CPU caches are categorized hierarchically
according to their affinity
to CPU cores.
Traditionally,
low-level caches (L1 and L2)
are private to individual CPU cores,
smaller in size and closer to the processor
and thus faster,
while the last-level cache (LLC)
is bigger in size and shared among cores.
When simultaneous multi-threading is enabled 
(\eg, hyperthreading in Intel),
two logical cores can share L1 and L2 caches.

CPU caches commonly adopt
the W-way set associative design,
where the cache memory is divided into
sets and each set holds W lines of
usually 64 bytes of data.
To locate desired data in a cache,
bits of a memory address
are divided into different sections --
offset to locate specific cache line,
index to determine the cache set, 
and tag to flag whether data is cached --
and utilized by addressing algorithms
to derive the exact location.
%
%When a process accesses data not present in a cache set,
%a cache \emph{miss} occurs and the target data
%will be fetched from main memory to
%the corresponding cache set.
%
%
%Note that we only consider L1 cache in this paper,
%where W is eight for modern Intel CPUs.

\subsection{Cache Attacks and Mitigations}
\label{ss:cache-attacks}
Cache attacks aim to leak information about
whether specific cache lines
have been accessed by a victim program.
Among various sources of leakage,
many of them utilize the time difference
between a cache hit and a cache miss.
Since a cache miss takes significantly longer
to retrieve the desired data,
the attacker can infer memory access patterns
of victim programs
by carefully manipulating cache lines mapped to the memory.

\PN{\primeprobe.}
\primeprobe attack and its variants ~\cite{prime-probe-dcache, prime-probe-icache, prime-abort}
monitor victim access to cache lines 
within a specific cache set throughout two phases.
The attacker keeps accessing the
target cache set so that it is completely
filled with the attacker's data (\emph{prime} phase).
Then,
the attacker waits for a
predetermined amount of time
and again accesses the data
he has loaded previously
while measuring the load latency
(\emph{probe} phase).
If the victim has accessed the target cache set,
some of the attacker-primed data will be evicted,
causing the reloading of the attacker's data
to take longer due to cache misses.
As a result,
the attacker can infer the victim's memory access activities
at a cache set granularity.
However, \primeprobe style attack incurs 
a significant amount of cache miss events, 
making the attack more easily detectable.

\PP{Preloading and cache pinning}
For decades,
researchers have been trying to
mitigate cache side-channel attacks in various ways
~\cite{random-fill-cache,pl-rp-cache,stealthmem,cachebar,catalyst}.
Recently,
~\cite{wait-minute, fine-grain-cross, evict+time, smokebomb, sbox-not-secure}
have discussed preloading sensitive data into the cache
to eliminate cache traces left by
the victim's data accesses.
Furthermore,
Gruss \etal~\cite{cloak}
propose Cloak, which utilizes \TSX
to preload and pin sensitive data and code
in transactional memory during execution.
If the attacker evicts corresponding cache lines,
\TSX allows the victim process to be immediately interrupted
and capture the malicious behavior.
Note that Cloak can provide stronger cache defense
compared to naive preloading which cannot prevent
attacker from interfering between prefetch and genuine accesses.
Cloak is also applied to \SGX to mitigate cache side-channel attacks
targeting enclave programs.
Notwithstanding the defenses posing a limit
to the level of leakage through the cache status,
\sys demonstrates that attacker can still leak information
through the underlying eviction policy
without causing evictions of the victim's cache lines.

\subsection{Cache Replacement Policies}
\label{ss:cache-policy}
If entire cache lines of one set are filled,
a next cache miss in that set will \emph{evict}
one of its present cache lines
in order to host the newly fetched data.
\emph{Cache replacement policy}
determines 
which cache line to evict.
%
%Since it takes significantly longer
%for the processor to fetch data from
%main memory than directly from the cache,
%an optimal cache replacement policy is
%crucial to processor performance.
%
In concept,
the replacement algorithm
keeps track of certain metadata
of the cache lines according to the
history of cache accesses,
on which the selection of the best cache way to evict
is based.
%
%As shown in~\autoref{f:prime},
%whenever one cache way is accessed
%due to either a cache hit or a cache miss,
%eviction metadata will be updated accordingly
%(tree shaped metadata). 
%
%Among all L1 cache replacement policies,
%the Least-Recently Used (LRU) policy and its variants
%are most widely adopted by modern CPUs 
%thanks to their high cache hit rates
%in common workloads.

\PP{LRU}
The Least-Recently Used cache replacement policy
is based on the assumption that
it is more likely for the processor to use
the more recently fetched or accessed data
than the stale ones.
To preserve the temporal locality,
the LRU algorithm keeps track of the age of each cache line
and chooses to evict the least recently used (oldest) cache way
upon cache misses.
%
%It takes $\log _{2}W$ bits to store the age of each cache line
%in a W-way set associative cache, 
%and $W\log _{2}W$ in total for each set.
%
Due to the expensive memory requirement and latency
in storing age information and updating LRU records,
an approximation of the LRU policy called
Pseudo Least-Recently Used (PLRU) is often used.
%to track a \emph{less}, instead of the \emph{least}, 
%recently used entry in 
%real-world implementation.

\begin{figure}[!h]
\centerline{\includegraphics[width=\columnwidth]{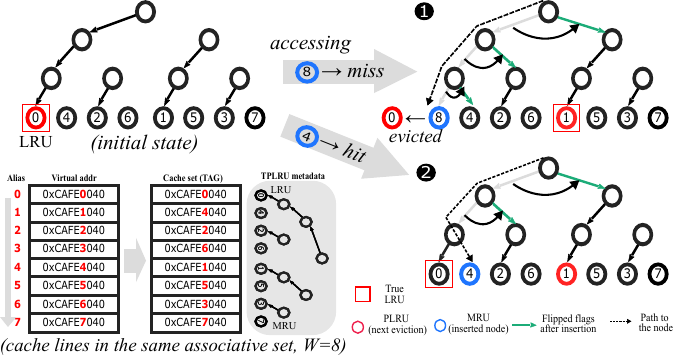}}
\caption{Tree-PLRU in action. It illustrates
  how the internal metadata of Tree-PLRU
  changes on a cache miss and hit on a cache line of the same associative set.
  % In \protect\BC{2},
  % Tree-PLRU considers cache way \protect\WC{1}
  % as the best LRU (PLRU) cache way,
  % unlike an ideal LRU policy would still consider cache way \protect\WC{0}
  % as the LRU cache way,
  % because of the tree-based metatdata that approximates the
  % PLRU ranks of cache ways.
}
\label{f:prime}
\end{figure}

\PP{Tree-PLRU}
The Tree-based Pseudo Least-Recently Used policy
(Tree-PLRU)~\cite{tplru} uses a binary tree structure
as its metadata
to encode the Pseudo Least-Recently Used (PLRU)
relationship within a cache set.
The cache lines in the same set
are divided recursively in binary
to sub-groups until it reaches 2-way groups.
As a result,
the division produces a binary tree (see~\autoref{f:prime}).
Each leaf node of the binary tree
represents the physical location of
a cache way (\cc{TAG}).
Each intermediate node
has a flag that indicates
the less recently accessed sub-group
represented as a sub-tree in the binary tree.
To find the PLRU cache way
that best approximates the LRU way,
Tree-PLRU starts from the root node,
and traverses along the less recently accessed sub-trees.
The finally reached leaf node then
contains the best approximated PLRU cache way
under Tree-PLRU.
In addition,
when a cache line is accessed,
all the flags on the reaching path
from the root node to the leaf node associated with the accessed line
are updated
to lead away from the path,
making the cache way the most recently accessed (MRU).
%
%With the Tree-PLRU algorithm,
%only $W-1$ bits are required,
%each of which represents the flag of a non-leaf node,
%% unlike an ideal LRU which needs $W\log _{2}W$ bits,
%to construct the binary tree for a W-way cache set.

\PP{Tree-PLRU in action}
Upon a cache hit or a cache miss,
Tree-PLRU updates its metadata
that represents the Pseudo-LRU ranks of cache lines
per associate set.
We define the \emph{PLRU rank} of a cache way
as the placement of the cache way
regarding the eviction order determined by the specific PLRU policy.
The sooner a cache way can get evicted under the specific PLRU policy,
the higher placement in the PLRU eviction order the cache way has
and thus,
the higher PLRU rank.
In the case of a cache miss (\BC{1} in~\autoref{f:prime}),
Tree-PLRU evicts the PLRU node
(\eg, \WC{0} in~\BC{1} reached by chasing the pointers from the root node)
and flips all the flags
along the path to lead away from the evicted node,
indicating the node as the most recently accessed one
(\ie, no pointers now directing to the node).
In the case of a cache hit (\BC{2} in~\autoref{f:prime}),
it simply flips all the flags
along the reaching path to lead away from the accessed cache line,
which similarly indicates it as the now most recently accessed one.
One significant side effect we leverage in our attack
is how the LRU status is approximated in Tree-PLRU as PLRU ranks.
By virtue of the approximation,
the LRU entry will not match
the entry marked as the highest PLRU rank
in some circumstances.
For example, in~\autoref{f:prime}-\BC{2},
Tree-PLRU considers \WC{1} as the next eviction entry (PLRU)
after \WC{4} is accessed due to the tree-based metatdata
that approximates the PLRU ranks of cache ways,
although \WC{0} is the true LRU entry.

\subsection{Intel TSX}
% NOTE. don't even mention lock elision, making people confused.
Intel Transactional Synchronization Extensions (TSX)
implements a hardware-based transactional memory
that allows multiple hardware threads to run critical sections concurrently
by detecting conflicts on shared data.
%
%To make sure
%both hardware threads run without conflicts
%(\eg, concurrent accesses to the same memory),
%TSX keeps track of
%two memory sets,
%\emph{reads} and \emph{writes},
%for each hardware thread.
%%
%When a transaction completes without any conflict,
%both sets will be made visible (\ie, committed)
%to the other threads.
%
When a transaction is aborted due to a conflict,
all monitored data will be discarded,
thereby restoring to the initial state
(\ie, rolled back).
%
% Upon the transaction abort,
% it is the software's responsibility
% to retry the same transaction
% in order to make a further progress.
%

%\PP{Intel SGX}
%Intel Software Guard Extensions (SGX) is a set of
%x86 ISA extensions that provides a hardware
%Trusted Execution Environment (TEE) called enclave,
%where code and data are protected
%from privileged attackers such as 
%operating system and hypervisor.
%%
%%Special memory regions exclusive to enclave usage are reserved
%%and encrypted by the CPU during program execution.
%%
%In this paper,
%we consider SGX as a restricted environment
%that poses synchronization challenges to the attacker (\autoref{ss:threat-model}).
%%
%We show that our attack succeeds with low noise
%under the SGX threat model (\autoref{ss:attack-sgx}, \autoref{s:eval}).

%\input{overview}
\section{Demystifying L1 Eviction Policy}
\label{s:reverse}
In this section,
we explain 
how we examine the details of
the undocumented L1 cache eviction policy
of modern Intel processors and Apple's M1 processor.
To understand the underlying cache architecture,
we use Intel TSX as a noise-free monitor
to capture specific L1 cache eviction events
on Intel processors.
For the same purpose on the M1 processor.
we use the undisclosed Apple's performance counter.
%
%Our interest in reversing L1 eviction policy lies in
%introducing novel methods that allow attackers
%to monitor internal cache architectures of Intel and Apple processors respectively.
Although we provide systematic approaches to 
collect cache traces, 
for recovering the policy from the trace,
we refer to \cite{cachequery, nanobench}.
%
%To the best of our knowledge, 
%this is the first work that provides detailed 
%analysis about the cache architecture of the M1 processor.

\PP{In-order memory accesses}
The internal states
of an eviction policy
can highly depend on the order of
cache accesses.
However,
modern processors adopts out-of-order execution 
and memory speculation.
To retain the expected order of cache accesses,
we craft in-order memory operations and time measurement primitives
by carefully inserting memory barrier instructions
and using pointer chasing~\cite{l1covert}.

\subsection{Intel Processor}
\label{ss:intel-reverse}
Reverse engineering
the cache eviction policy
is not a new problem in the x86 architecture
\cite{reload+, nanobench}.
However,
achieving it without noise 
in user space is challenging.
For example,
Briongos \etal~\cite{reload+}
encounters false positives
due to the adoption of timing based measurements,
while nanoBench~\cite{nanobench} leverages
hardware Performance Monitoring Extension (PMU)
that requires root privilege to achieve accurate results,
restricting its scope of application.
We devise a user-space noise-free
reverse engineering technique utilizing \TSX,
and uncover that the underlying eviction policy
is Tree-PLRU.
We used the same environment setting described in \autoref{s:eval}.
%
%The key intuition
%of our noise-free reverse engineering technique
%is to leverage Intel TSX and
%its explicit event (\ie, TSX abort) delivered to user space software
%in verifying the eviction policy of Tree-PLRU,
%\ie, which cache line is evicted in what order
%from the associative set.

%Memory addresses read from within a transactional region constitute the read-set of the transactional region and addresses written to within the transactional region constitute the write-set of the transactional region
\PP{Different behaviors of TSX read-set and write-set}
\TSX, within a transactional region,
tracks memory addresses written into as a \cc{write-set} and
memory addresses read from as a \cc{read-set}.
To prevent conflicting accesses
where another logical processor either reads memory from the write-set or 
writes to memory of either the read- or write-set,
\TSX monitors such events at different cache levels. 
When a cache line mapped to an address in the write-set or the read-set 
is evicted from the L1 data cache or the L3 cache, respectively,  
it triggers TSX abort resulted from the conflicting access.

\PP{Algorithm for reversing the L1 cache with TSX}
\SetCommentSty{commfont}
\SetKwRepeat{Do}{do}{while}
\SetKwInput{KwInput}{Input}
\SetKwInput{KwOutput}{Output} 
\SetAlFnt{\footnotesize}
\begin{algorithm}[h]
\DontPrintSemicolon

	\KwInput{X[nWays], Y[nWays]: Addresses in set X and Y are mapped to the same set, but X and Y are disjoint.}
  \KwOutput{evictSeq[nWays]: Num of required evictions per cache line}
  \For{$target\leftarrow 0$ \KwTo nWays}{
    \Do{TSX abort} {
      \For{$numEvict\leftarrow 1$ \KwTo nWays+1}{
        \emph{beginTSX}\;
        \For{$i\leftarrow 0$ \KwTo target}{
          memRead(X[i])\;
        }
        \emph{memWrite(X[target])}\;
        \For{$j\leftarrow target+1$ \KwTo nWays}{
          memRead(X[j])\;
        }
	\emph{memRead(X[read_primed])}\;
        \For{$k\leftarrow 0$ \KwTo numEvict}{
          memRead(Y[k])\;
        }
        \emph{endTSX}\;
      }
    }
	\If{TSX aborts} {
  	  evictSeq[target] = numEvict\;
    	}
  }

\caption{TSX-supported eviction policy reversing}

\label{alg:rev}
\end{algorithm}

The essence of TSX-assisted reversing is utilizing
different condition of conflicting access on the read-set and the write-set.
Regardless of memory write or read operations, 
the cache lines are fetched to the L1 data cache. 
Nevertheless, a TSX abort will be triggered 
only when a cache line in the write set is evicted 
from the L1 cache.
Note that an abort will not be triggered 
when a cache line in the read-set is evicted from the L1 or L2 cache.
Therefore, we can reliably measure 
how many fresh cache misses, thus evictions,
are required to evict a specific cache line tracked by the write-set
from the L1 data cache.
Note that this number of eviction represents
the internal decisions of the cache replacement policy.
%
%In detail, 
%we locate a cache line under monitoring in a write set
%and the rest of the cache lines in a read set of \TSX.
%In our~\autoref{alg:rev},
%we examine
%how many cache misses, thus evictions,
%are required (\cc{numEvict})
%to evict each cache line in the same associative set
%(\cc{target}).
%
In each round of~\autoref{alg:rev},
we load the \cc{target - 1} number of the cache lines (Line 5-6) to the read-set,
the target cache line to the write-set (Line 7),
and the rest of the cache line to the read-set (Line 8-9).
At this stage,
since the associative set contains all the data (from \cc{set X}) we loaded, 
an additional cache line accesses to the same associative set
will incur an eviction (Line 11-12).
Then,
we count how many \emph{new} cache lines (from \cc{set Y})
can be loaded
until the target cache line is chosen for eviction
by the underling eviction policy.
Note that
any L1 data cache eviction from the read-set would not cause
the TSX to abort.
%
%With this algorithm,
%we record how many misses are required for
%evicting each cache line in the same associative set.
%
%
Furthermore,
to understand how the underlying eviction policy is affected by the L1 cache hit events, 
we enumerate different combinations of read operations on the primed data (Line 10).
Therefore,
we expose the
eviction orders of the cache lines
within the same associative set
resulting from all combinations of cache hits
and learn that the changing eviction order conforms to the behavior of Tree-PLRU.

%\PP{Reversing in action}
%We depict how to use this general algorithm to
%reverse engineer the L1 cache eviction policy
%in~\autoref{f:prime}.
%%
%As shown in the table,
%eight virtual addresses,
%\cc{0xCAFE0040} to \cc{0xCAFE7040} (array \cc{X}),
%are use to prime to a particular cache set.
%%
%To record how many evictions (\cc{numEvict}) are required
%to evict \cc{0xCAFE0040} (\WC{0}),
%we place it into the cache with write operations (TSX write set)
%and others with read operations (TSX read set).
%%
%When there is no further access to the primed set (\BC{1}),
%accessing one element in the array \cc{Y}
%will evict \cc{0xCAFE0040} and raise a TSX abort,
%making \cc{numEvict} equal to \cc{1}.
%%
%However,
%when a cache hit,
%on \cc{0xCAFE4040} (\WC{4}) for example,
%happens to the primed cache set
%before any cache miss caused by array \cc{Y} (\BC{2}),
%eviction order will alter in the way
%following the underlying implementation of the PLRU policy,
%and a number of evictions not following the ideal LRU can be recorded,
%\ie, making \cc{numEvict} equal to \cc{4} instead.
%%
%By enumerating different combinations of cache hits on the primed data
%before introducing evictions,
%we are able to record different sequences of
%required numbers of evictions.
%%
%Therefore,
%we expose the
%eviction orders of the cache lines
%within the same associative set
%resulting from all combinations of cache hits
%and learn how the eviction order changes
%conforms to the behavior of Tree-PLRU.

\subsection{Apple M1 Processor}
\label{s:m1-reversing}
It is challenging to explore 
the microarchitectural behaviors
of the newly released Apple M1 processor,
as its internal states and interfaces are not previously studied.
Nonetheless, we successfully reveal that 
the M1 processor adopts the Tree-PLRU eviction policy 
for the L1 data cache.
To reach the obtained results, 
we utilized the Apple's hardware PMU 
to monitor its L1 data cache events
because the M1 processor is not equipped with transactional memory. 
In detail, 
we utilize the fact that 
L1 cache hit and miss events have different latency.
We expect that TSX-style reverse engineering (\autoref{ss:intel-reverse}) is possible 
with Transactional Memory Extension~\cite{arm:tme}
deployed in ARMv9~\cite{armv9}.
We used the same environment setting described in \autoref{s:eval-m1}.

\PP{Reversing with the undocumented PMU}
The first challenge we faced in reversing the M1 architecture is 
the lack of a publicly available interface as a measurement tool,
such as hardware PMUs including a high-resolution timer.
Although the M1 architecture expands the ARMv8,  
we found that the M1 does not deploy the standard ARMv8 PMU~\cite{armv8}
(\ie, reading and writing system registers of ARMv8 PMU halts the system).
Therefore, we reverse-engineered the 
interface of the Apple's proprietary PMU and related events.
Although no official document from Apple discloses such information, 
we found that Apple's XNU kernel partially utilizes their proprietary PMU.
Based on the analysis and experiment,
we found that the M1 processor can concurrently monitor
at most 10 PMU events per core, 
and each PMU can track one of the 255 undocumented events (0 to 254).
We also empirically found two undocumented PMU events: 
\emph{0x2} that tracks the core clock cycles (\eg, \emph{rdtsc} as in x86)
and \emph{0xa3} that monitors the L1 data cache miss event.
Those two PMU events are enough to uncover 
the replacement policy of the L1 data cache.

\PP{Cache hierarchies of M1}
Information about cache hierarchies such as 
set associativity, number of sets, and cache line size 
is essential for reversing the replacement policy.
Unfortunately, to the best of our knowledge, 
there is no public documentation specifying 
the M1 processor's cache hierarchies.
Therefore, we for the first time retrieve such information 
by utilizing the system registers \emph{CCSIDR_EL1} and \emph{CSSELR_EL1}.
As the M1 architecture adopts two different types of cores (Firestorm and Icestorm),
we collect two sets of cache hierarchies respectively. 
All the detailed information is described in \autoref{t:m1-data-cache}.
 \begin{table}[h]
   \centering
   \resizebox{\columnwidth}{!}{%
\begin{tabular}{lcccc}
\toprule
\multirow{2}{*}{CPU Type} & \multicolumn{2}{c}{FireStorm} & \multicolumn{2}{c}{IceStorm}
\\\cmidrule(lr){2-3}\cmidrule(lr){4-5}
& L1 DCache  & L2 DCache & L1 DCache & L2 DCache\\\midrule
Number of Set           & 256        & 8192      & 128     & 2048       \\
Set Associativity       & 8          & 12        & 8       & 16         \\
Cache Line Size(Byte)   & 64         & 128       & 64      & 128        \\\bottomrule
\end{tabular}
}

%M1 big core
%L1 cache
%Cache set: 256,  Cache associativity: 8,   Cache Line Size: 64B
%L2 cache
%Cache set:8192, Cache associativity: 12, Cache Line Size:128B
%
%M1 small core =============
%L1 cache
%Cache set:128,   Cache associativity: 8,   Cache Line Size: 64B
%L2 cache
%Cache set:2048, Cache associativity:16   Cache Line Size: 128B

         \caption{Information of data cache hierarchies in M1 processors}
   \label{t:m1-data-cache}
 \end{table}

\PP{L1 data cache set mapping}
To measure the L1 data cache activities in a particular set,
we should be able to manipulate the addresses mapped to a particular set.
For example, 
we need at least 9 addresses mapped to a particular set
as the L1 data cache is 8-way set associative in both types of the cores.  
We found that M1 employs different 
cache set mappings depending on the core type.
For all virtual address $x$, 
its L1 data cache set $s$ 
is determined by the following equations:
$s = (x \bmod 16384) / 64$ on Firestorm core and 
$s = (x \bmod 8192) / 64$ on Icestorm.

\begin{figure}[!h]
\centerline{\includegraphics[width=\columnwidth]{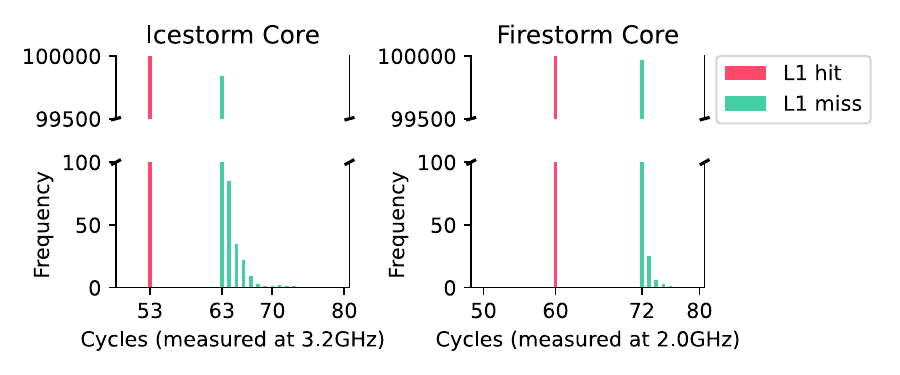}}
\caption
{
	L1 hit and miss latency for Icestorm and Firestorm core.
	The measured latency includes the memory barrier instruction's latency: 51 cycles and 56 cycles, respectively, on Icestorm and Firestorm core.
	Therefore, on two different cores, 
	the actual L1 hit latency is 2 cycles and 4 cycles,
	and L1 miss latency is around 12 cycles and 16 cycles, respectively.
	Note that each core runs at different clock frequency.
}
\label{f:m1-l1-latency}
\end{figure}

\PP{L1 cache hit and miss latency}
To measure the L1 hit latency, 
we first prefetch one memory address with \cc{ldr} instruction 
and then measure the latency of loading the same entry once again. 
If the entry has been prefetched before being accessed in the measurement,
it will always result in an L1 hit.
%
%In detail,
%we read twice the PMC register tracking the core clock cycle, 
%before and after the \cc{ldr} instruction,
%and calculate the elapsed core clock cycles for executing the \cc{ldr} instruction.
%
Although we have no information about the eviction policy yet,
we can still measure the L1 miss latency because 
we are aware that each L1 data set consists of 8 cache lines (\autoref{t:m1-data-cache}). 
In detail, 
we access 16 memory addresses mapped to a particular L1 data set 
and measure the latency to access the first entry once more.
Note that the first 8 entries are evicted from L1 caches
after the following 8 accesses occur. 
Therefore, the accesses to the first 8 entries 
will not be served by the L1 data cache.
As shown in~\autoref{f:m1-l1-latency},
we can clearly distinguish L1 hits from L1 misses in
both types of the cores. 
We implement the reversing logic as a kernel driver
and launch measurements on a designated processor
using \cc{smp_call_function_single}
to eliminate the noise introduced by context switching.

\PP{Reversing M1 eviction policy}
To reverse engineer the eviction policy of the L1 cache on the M1 platform,
we follow the similar reversing logic described in~\autoref{alg:rev}.
Based on the latency difference measured in~\autoref{f:m1-l1-latency},
we can determine which cache line has been evicted and
recover the eviction policy from the trace.
Also, 
we utilize the \cc{0xa3} PMU events together 
to monitor L1 cache activities
for better accuracy. 
Due to the lack of space, implementation details
are described in~\autoref{ss:m1-reverse}.

\section{\sys}
\label{s:attack}
\sys has three distinctive characteristics
compared with known cache side-channel attacks
(see~\autoref{ss:cache-attacks}).

\noindent
\textbf{\WC{1} Stealthy.}
\sys does not incur eviction of the victim's prefetched data
to leak the access patterns.
Since \sys only accesses the cache lines primed by the attacker
to manipulate eviction metadata
(\ie, internal states of Tree-PLRU),
it does not interfere with
the victim's execution
or destruct cache state.
This property is particularly important
because existing defenses against
\primeprobe and \flushreload
rely on preloading and locking
the cache sets~\cite{cloak}
to prevent cache evictions.

\noindent
\textbf{\WC{2} Minimal synchronization.}
When an attack requires multiple cache accesses, 
it is more challenging to synchronize those accesses with the victim's access.
Thus, incorrect synchronization incurs noise and makes the attacks unreliable. 
Furthermore, when a prefetch-based defense is deployed, 
especially for \primeprobe style attacks,
the targeted windows tend to be narrower 
because the attacker should evict the prefetched entries before they are consumed.
As a result, the attacker has to average out the noise by
repeatedly launching the procedure a significant number of times~\cite{reload+, prime-abort, cloak},
thus leaving a detectable level of microarchitectural traces for
detection-based countermeasures~\cite{cacheshield, cloudradar}.
In contrast,
\sys requires only one synchronized memory access to
the attacker's pre-primed cache lines,
minimizing both noise from synchronization efforts and
microarchitectural traces left.

\noindent
\textbf{\WC{3} Leakage via non-shared memory.}
Attacks such as
\flushreload and \reloadrefresh~\cite{reload+}
require shared memory between the attacker and the victim
to introduce changes on cache status, such as flushing the victim's cache line.
However,
the requirement highly limits the range of 
application scenarios.
%
%In contrast,
%\sys does not require any shared knowledge
%about the victim
%while still achieving address precision.
In contrast,
\sys does not require any shared memory resource
to leak the victim's access pattern.
Since the \sys attacker has complete knowledge about
the shared eviction policy,
the attacker can precisely reveal access patterns
of a target cache line
through the eviction metadata instead of the cache itself. 
%as long as a spy process can be co-located
%on the hyperthreaded core of the victim.

\begin{figure*}[!t]
\centerline{\includegraphics[width=1.00\textwidth]{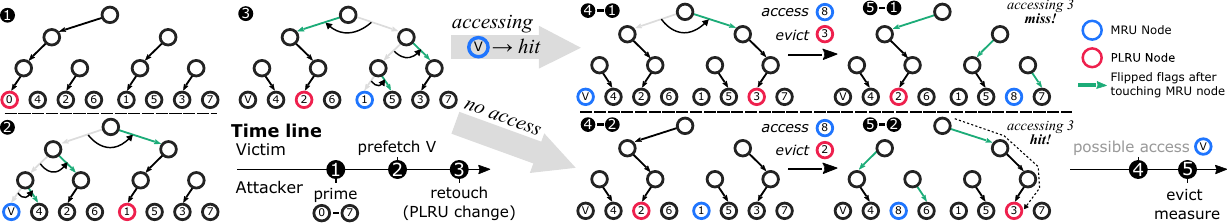}}
\caption
{
  An overview of \sys against the L1 cache with the Tree-PLRU policy. 
  Changes on Tree-PLRU are demonstrated following the timeline. 
  Depending on the presence of the victim's access following the prefetching,
  two different PLRU cache ways will be produced.
}
\label{f:overview}
\end{figure*}

\subsection{Attack Model}
\label{ss:threat-model}
\sys is a general attack applicable to any architecture meeting the below assumptions.
Note that further details will be described in~\autoref{s:eval},~\autoref{s:eval-m1}
based on the target architecture.
%
%\PP{Execution environment}
Regarding the execution environment, 
we assume that the L1 data cache is shared between attacker and victim.
We assume that the victim process prefetches data before accessing
to conceal the access pattern.
We also assume that cache evictions of sensitive data are not allowed.
%\PP{Attacker's capability}
Considering the attacker's capability,
we assume that 
the attacker can replay the security-sensitive code
unlimited times without introducing unexpected interruptions
to the victim process.
The attacker is not expected to
have any shared memory with the victim.
Lastly,
we assume that the attacker can
freely allocate virtual memory
to pick addresses that are mapped to 
desired L1 cache sets. 

\subsection{Leaky Tree-PLRU}
\label{ss:leaky-tplru}
The Tree-PLRU cache replacement policy
aims to mimic a true LRU policy
by using a tree-based data structure,
as explained in~\autoref{ss:cache-policy}.
One of the most distinct properties of Tree-PLRU is that
the eviction metadata used to decide the eviction order 
is encoded as a binary tree structure
when entries are placed into the cache.
Therefore, 
the exact eviction order for LRU is not accurately tracked
and approximated by sub-trees (Pseudo-LRU).
In detail,
one cache way is grouped 
with other cache ways
by sub-trees represented in the eviction metadata, and
the PLRU rank of a cache line (\ie, eviction order) can be affected by
where an access to or an eviction of cache lines in the same sets
occurs in the tree structure.
%
% \XXX{too sudden. you should clarify what's ``sub'' tree, etc.
%   the key property is that the eviction order
%   is not accurately tracked for LRU,
%   but is approximated as part the tree structure.
%   It means the eviction order of a cacheline
%   can be affected by
%   where an access to or an eviction of cachlines in the same sets
%   happens in the tree structure.
%   then, elaborate like below.
% }
%
We found that such a property can be
carefully manipulated by an attacker 
to produce eviction metadata
that leak the cache activities of the victim
without causing evictions.
We now explain how this can be achieved.

% \PP{Prerequisites}
% As the victim process is assumed
% to be protected by \FS\{transaction based?}
% cache prefetching and pining mitigations,
% no eviction is allowed during the protected period of execution.
% %

\PP{Indistinguishable back-to-back accesses}
Consecutive accesses to one cache line back-to-back
are observed as single memory access occurring under Tree-PLRU
since the Tree-PLRU approximates the order of accesses 
as a binary tree.
As we recall from~\autoref{ss:cache-policy},
when a cache line is accessed,
all the flags on the reaching path 
from the root node to the associated leaf node are updated
to lead away from the path,
making the cache line the most recently accessed (MRU).
However,
if the cache line is accessed again back-to-back,
the corresponding flags will then remain unchanged,
as they are already updated and indicating
the correct less recently accessed sub-trees.
As a result,
the fact that one cache line
has been accessed multiple times
cannot be revealed directly from
the final tree state
if the accesses occur back-to-back.
In the case of \sys,
since a prefetch operation
essentially accesses the prefetched data,
we cannot discover
a back-to-back victim access to the prefetched data
directly from the final tree state either.

\PP{Distinguish victim's access from prefetch operation by retouching}
To distinguish back-to-back accesses
to the same cache line solely from final tree states,
the attacker should be able to map one access to
exactly one time of observable tree state change.
Therefore,
by checking whether an additional change
in the tree state has occurred,
the attacker can discover the victim's access following the prefetch.
%\JH{what is the PLRU cache way change, no descriptions here!}
We define an observable tree state change
as a change of the node indicating the PLRU cache way
(\ie, next eviction entry).
Note that a change of the PLRU cache way is observed
only when the flag of the root node is flipped.
%
% To force another observable tree state change
% from the back-to-back victim access,
% the solution is to
% make the prefetched entry a part of the
% less recently accessed sub-tree again
% so that a following access will
% cause corresponding flags to change once more.
%
In~\autoref{f:overview},
attacker primes all cache lines of one set (\BC{1}),
and waits until the victim prefetches \WC{V} (\BC{2}).
%and allow the victim to prefetch \WC{V} (\BC{2}).
%
%To distinguish whether the victim accesses the prefetched data
%after the victim's prefetch operation, 
To discover the victim’s access following the prefetch,
attacker retouches the PLRU cache way (\WC{1}).
%produced after the victim's prefetch.
%
Consequently,
it flips the corresponding flags,
including the root node flag,
and produce a new PLRU node (\WC{2}),
causing the root node flag to direct toward
the prefetched entry (\BC{3}).
If a subsequent victim access
to the prefetched entry occurs (\BC{4}-\BC{1}),
the root node flag will be flipped again
to lead away from the prefetched entry,
announcing a new PLRU cache way (\WC{3}).
In contrast, 
\WC{2} remains as a PLRU cache way 
when there is no victim access after preloading (\BC{4}-\BC{2}).
If the attacker had not retouched the PLRU cache way,
both \BC{4}-\BC{1} and \BC{4}-\BC{2}
would have produced the identical PLRU tree,
which prevents discovering the victim's access following the prefetch.

\PP{Probe the latest PLRU cache way and match with cache activities}
When another miss event is introduced to the set,
PLRU cache way will be evicted and allow a new entry to be cached. 
However, note that different cache ways will be evicted
depending on which entry in the set is the PLRU entry,
which is determined by the possible victim access following the retouch.
We produce a cache miss by accessing \WC{8},
a new cache line associated with the current set,
to evict the current PLRU cache line in both cases.
Note that the evicted cache line belongs to attacker's primed set
and does not affect the victim's cache way.
Consequently,
different entries will reside in the set
(\WC{2} remain in \BC{5}-\BC{1}, and \WC{3} remain in \BC{5}-\BC{2}).
%
%Since the PLRU cache way
%that ought to result from \BC{4}-\BC{1} is \WC{3},
We can discover whether there were 
victim's access following the prefetch
by measuring the access latency of \WC{3}.
A cache miss latency indicates that \WC{3}
was indeed evicted and confirms that
the victim has accessed the prefetched entry
while a L1 cache hit latency indicates that
\WC{2} has been evicted instead,
indicating that there was no further access from victim.

\subsection{Synchronization in \sys}
\label{ss:assisted-sync}
 \begin{table}[h]
   \centering
   \renewcommand{\arraystretch}{1.1}
\resizebox{\columnwidth}{!}{%

\begin{tabular}{l c c c c c}
\Xhline{2\arrayrulewidth}
                   &$ \WC{A} P \rightarrow R \rightarrow A $ &
                    $ \WC{B} P \rightarrow R               $ &
                    $ \WC{C} P \rightarrow A \rightarrow R $ & 
                    $ \WC{D} R \rightarrow P \rightarrow A $ &
                    $ \WC{E} R \rightarrow P               $ 
\\ \hline
Retouch \WC{1}           & PLRU:3                    & PLRU:2          & PLRU:2              & PLRU:3              & PLRU:3      \\
PLRU Aware      & \textbf{PLRU:3}              & PLRU:2          & PLRU:2              & PLRU:2              & PLRU:2      \\
\Xhline{2\arrayrulewidth}
\end{tabular}

}

   \caption{PLRU cache ways
   	produced by all five possible operation sequences.
  	After applying PLRU Aware \retouchonly (\autoref{ss:post-sync},
  	the sequence \protect\WC{A} can be distinguished among 
  	all five sequences.}
   \label{t:five-case}
 \end{table}
In the previous section \autoref{ss:leaky-tplru},
we describe the attack only when
the attacker's retouch happens 
after the victim's prefetch (\WC{A} and \WC{B} in ~\autoref{t:five-case}).
However,
as shown in \autoref{t:five-case},
five different operation sequences can happen
depending on the order of the victim's and attacker's memory accesses
during \sys attack (\ie, synchronization).
In this section, we will demonstrate the challenges 
imposed by synchronization in \sys.
%and discuss the limitations of \emph{assisted synchronization}.

%Although this assumption temporarily alleviates the synchronization problem 
%in \sys, we will continuously adopt this assumption
%until ~\autoref{s:impl}
%to showcase the essence of \sys, 
%leaking victim's access from eviction policy.
%%
%In this section we continuously assumes that 
%attacker retouch happens synchronously after the victim's preloading
%with the help of assisted synchronization technique 
%such as ~\cite{prime-abort}.
%After then, we will describe 
%challenges of \sys and 
%limitations of assisted synchronization.

\begin{figure}[!t]
\centerline{\includegraphics[width=\columnwidth]{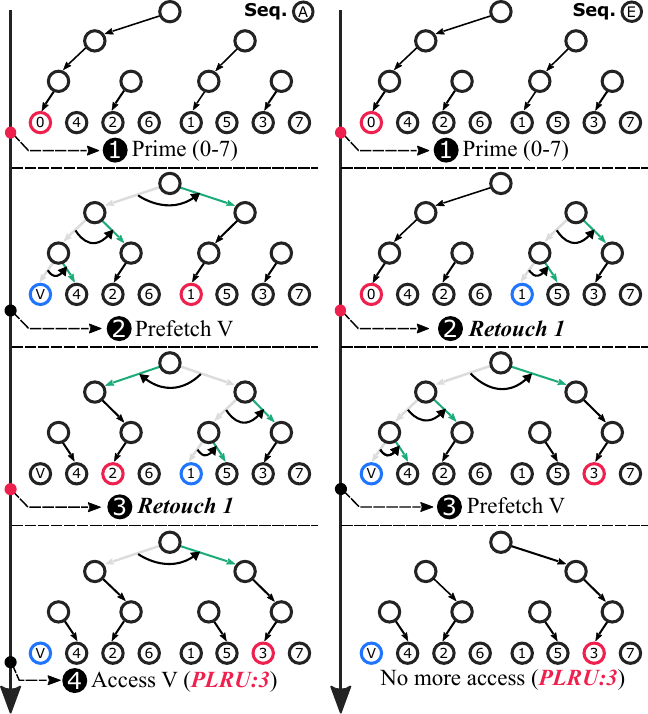}}
\caption
{
	When \retouchonly \protect\WC{1} is applied,
	the identical Tree-PLRU metadata is produced 
	as a result of two different operation sequences: 
	synchronized (left) and unsynchronized (right). 
}
\label{f:false-positive}
\end{figure}

\PN{Retouching \WC{1} can produce false positive results
of distinguished victim accesses.}
If the attacker cannot adequately achieve precise synchronizations,
unwanted operation sequences are produced and
result in noisy Tree-PLRU metadata,
posing \emph{false positives}.
Particularly,
when an attacker's retouch
occurs prior to the victim's prefetch,
two extra operation sequences \WC{D} and \WC{E} (we call them unsynchronized sequences) 
can occur.
%can occur other than the three introduced in~\autoref{ss:leaky-tplru} (\ie, \WC{A}, \WC{B}, \WC{C}):
%
\begin{flalign*}
&\WC{D} Retouch_{Attacker} \rightarrow Prefetch_{Victim} \rightarrow Access_{Victim} \\
&\WC{E} Retouch_{Attacker} \rightarrow Prefetch_{Victim}
\end{flalign*}
As shown in~\autoref{f:false-positive},
although synchronized sequence \WC{A} and
unsynchronized sequence \WC{E}
represent opposite victim cache activities (access vs. no access),
they result in the same Tree-PLRU eviction metadata.
Note that unsynchronized sequence \WC{D} results in the same Tree-PLRU 
metadata as \WC{E} because 
the following victim's access occurs immediately after the prefetch.
In fact,
the mixture of all five possible sequences
makes the previously unique PLRU cache way
produced by the desired sequence \WC{A} completely
indistinguishable, as shown in~\autoref{t:five-case}.
Therefore,
without techniques to
differentiate such conflicting cases,
final tree states resulting from
unsynchronized retouch
cannot render useful information
about victim cache activities.

%The ultimate goal of the attacker is 
%exclusively distinguishing the sequence \WC{A} and \WC{B} from others
%to understand whether the victim access has been made after the preloading. 
%%
%The \sys described in \autoref{ss:leaky-tplru} can 
%uniquely differentiate between \WC{A} and \WC{B}
%by retouching \WC{1} after the victim's prefetch.
%However, when the retouching happens earlier than the prefetching,
%we call it unsynchronized reotuching, 
%the sequence \WC{A} can not be distnguished from others (\WC{D}, \WC{E}).
%because all three sequences result in same Tree-PLRU state.
%The reason of \emph{false positives} due to the unsynchronized retouching 
%will be elaborated on later in~\autoref{ss:unsync}.

\PP{Synchronization can eliminate false positives}
The root cause of false positives is 
unsynchronized retouching before the victim's prefetching.
Therefore, 
if the attacker is able to
successfully retouch \WC{1} after the victim entry has been prefetched,
only three possible operation sequences
can occur and eliminate the false positives:
\begin{flalign*} 
&\WC{A} Prefetch_{Victim} \rightarrow Retouch_{Attacker} \rightarrow Access_{Victim} \\
&\WC{B} Prefetch_{Victim} \rightarrow Retouch_{Attacker} \\
&\WC{C} Prefetch_{Victim} \rightarrow Access_{Victim} \rightarrow Retouch_{Attacker}
\end{flalign*}

Each sequence results in
different tree states following Tree-PLRU policy,
where the PLRU cache way resulting from
sequence \WC{A} is different from the other two (\WC{B}, \WC{C})
Since we know the sequence \WC{A} produces
a unique PLRU cache way among the three,
we can probe whether sequence \WC{A} has occurred
by forcing a cache miss and check whether
indeed the unique PLRU cache line got evicted.

\section{Pseudo Synchronized \sys}
\label{s:impl}
% %For the previous side channels that has been no false positives introduced by the attacker's access
% In most of the previous side channels,
% when there is no noise 
% caused by unexpected events such as interrupts and context switching
% there are only two cases that attacker can encounter:
% capture the victim's access or miss it. 
In this section,
we introduce the pseudo synchronized \sys attack, 
which does not require precise synchronization
but still achieves the same leakage
by purely relying on Tree-PLRU metadata.
We first look at
characteristics of Tree-PLRU metadata 
that create \emph{false positives}
in distinguishing the victim access from prefetch
when assistance for precise synchronization
is not available.
We then propose a new technique
called \emph{PLRU Aware \retouchonly},
which cleverly solves the challenge 
introduced by unsynchronized \retouchonly
and shows that when applied,
the attacker can achieve the same leakage
as with precise synchronization.

\subsection{Limited Precise Synchronization Techniques}
\label{ss:limited-sync}
Various synchronization techniques to 
timely introduce an attacker's interference
have been discussed under various threat models~\cite{prime-abort, sgx-step}.
Nonetheless,
it is not always the case that such assistance is available 
to achieve precise synchronization.

\PP{Hardware transactions.}
\primeabort~\cite{prime-abort} allows the attacker
to accurately observe a victim's cache access using TSX abort.
%
%Attackers will first initiate
%a TSX transaction before the \primeonly phase.
%%
%Once the \primeonly phase finishes,
%the attacker simply waits for a TSX abort,
%which precisely indicates that the victim process has
%accessed the target cache set.
%%
Intuitively,
a \sys attacker might also rely on such a technique
to precisely synchronize retouch with
the victim's prefetch.
Unfortunately, TSX-assisted synchronization
could not work in our threat model.
%
%when the victim process runs inside SGX enclaves
%on hyperthreaded cores with the latest security updates,
%the \primeabort style precise synchronization technique is not feasible.
%
From the latest stepping R0~\cite{hw-patch,taa},
Intel has deployed hardware changes
as mitigations against microarchitectural side channels~\cite{zombieload, ridl, taa, spectre, meltdown}.
This mitigation prevents 
a hyperthreaded core from handling TSX transactions
while its sibling core runs an SGX enclave process.
Consequently,
even if the co-located attacker initiates a TSX transaction
before the victim executes,
the transaction will immediately abort
upon the victim's entering the enclave
without capturing any cache activity. 
Furthermore, 
TSX is disabled by default in modern operating system;
and user-level attacker in \sys is prohibited from utilizing it.

%\PP{Precise interrupts}
%Recent research~\cite{sgx-step, high-res,nemesis,cachezoom}
%has found that 
%precise interrupts raised by the Intel APIC timer 
%allows attacker to infer execution states of the enclaves.
%%
%In contrast,
%any interrupt raised during a TSX transaction
%will cause a TSX abort.
%%
%When an interrupt is raised inside a TSX transaction,
%the TSX abort handler will be invoked before 
%the interrupt handler.
%%
%Therefore,
%a \sys attacker in SGX will not be able to
%use the APIC timer to
%single-step the SGX enclave execution
%protected by TSX~\cite{cloak}
%and achieve precise synchronization.
% \FS{necessary to mention SGX here?}

\begin{figure}[!t]
\centerline{\includegraphics[width=\columnwidth]{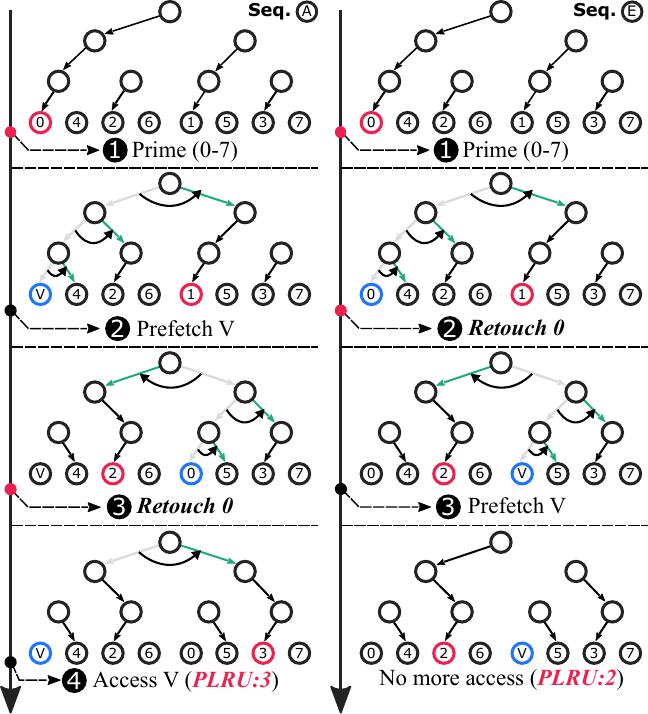}}
\caption
{
   When PLRU Aware \retouchonly (\autoref{ss:post-sync}) is applied
   to the case in ~\autoref{f:false-positive},
   by probing the PLRU cache way
   represented in the final tree states,
   the attacker is able to distinguish 
   the two operation sequences.
}
\label{f:equalizing}
\end{figure}
\subsection{Unsynchronized \retouchonly}
\label{ss:unsync}

\PP{The attacker's dilemma}
% This characteristic results from limited capability of eviction tree
% with respect to tracing access orders of cache entries in one set. 
% Compared to true LRU eviction policy 
% which can record complete access sequences with separate counters,
% eviction tree in TPLRU policy can record partial access order
% because all accesses are accumulated in single eviction tree.
Here comes the attacker's dilemma:
%as we proposed in~\autoref{ss:leaky-tplru},
The PLRU cache way produced 
after the victim's prefetch and attacker's retouch operation, 
should belong to the same sub-tree of the victim's cache line 
(root-to-left or root-to-right sub-tree); however, 
different orders of the two operations make 
the PLRU cache way to be located in different sub-tree.
The fact that the PLRU cache way resides in 
the same sub-tree as the victim's cache line 
guarantees that the victim's access following the prefetch operation
to be traced in the Tree-PLRU structure. 
Therefore, 
at the time of the two operations finished, 
if the victim's entry and PLRU cache way
are not located in the same sub-tree,
\sys can misinterpret the Tree-PLRU structure. 

As depicted on the right side of \autoref{f:false-positive},
the attacker unexpectedly retouches \WC{1}
when it is not synchronized with the victim's prefetch operation 
(sequence \WC{E}).
The accessed cache line does not reside in the sub-tree
that contains the PLRU cache way (\ie, root-to-left sub-tree), so 
the PLRU cache way under Tree-PLRU
does not change.
However,
the PLRU ranks of other cache lines
within the opposite sub-tree (\ie, root-to-right sub-tree) 
has been changed 
(\eg, promoting sub-tree containing \WC{3} and 
demoting sub-tree containing \WC{1}).
Consequently,
the effect on the eviction candidate is preserved
% preserved because more recently accessed nodes are 
% not on the reaching path of any less recently accessed nodes,
% isolated by the root node,
% thus will not be affected by eviction
and postponed until the next root node flip,
\ie, the next access to the less recently accessed sub-tree.
As a result of the following victim's prefetch operation (\BC{3}),
PLRU rank changes captured in the Tree-PLRU structure, which was incurred by \BC{2}, 
makes the \WC{3} as the PLRU cache way. 
On the other hand,
the synchronized (left) attacker retouches \WC{1} when it is in
the PLRU cache line after prefetch (\BC{2}),
causing \WC{3} to have the
highest plru rank in the sub-tree (\BC{3}).
As a result,
both the following victim access
when synchronized 
and the victim prefetch when unsynchronized (\BC{3})
flip the root flag and produce \WC{3} as the next PLRU cache way.
To solve the dilemma
under unsynchronized settings,
retouch needs to do more than just
mimic an access.

\subsection{PLRU Aware \retouchonly}
\label{ss:post-sync}
Recall that the root cause of the attacker's dilemma is the
nondeterministic location of the PLRU cache way
represented in the Tree-PLRU structure when not synchronized.
The attacker wishes to always retouch the PLRU cache line
to cause an immediate observable tree state change,
but might unexpectedly retouch a cache way with low PLRU rank (high MRU rank)
prior to victim prefetch and cause an indistinguishable final tree state.
Our solution is called PLRU Aware \retouchonly,
a special type of \retouchonly that always accesses
the PLRU cache line.
In order to always retouch the PLRU cache way,
the attacker should either evict the cache line in
the PLRU cache way,
or directly access the PLRU cache way.
PLRU Aware \retouchonly solves the dilemma
by always retouching a cache line that satisfied both conditions,
the resulting PLRU cache line immediately after the attacker prime.
Note that the attacker knows
the resulted PLRU cache line,
as he is aware of the eviction policy.
The reason behind is that
retouching such a cache line will either bring back the cache line
to the PLRU cache way after being evicted by victim prefetch
by a cache miss caused by the following attacker retouch (left side in \autoref{f:equalizing}),
or access the PLRU cache way itself before victim prefetch (right side in \autoref{f:equalizing}).
% \FS{maybe a fig?}
% \FS{comparison of the effect caused by
% a cache miss and access eviction entry is shown in figXXX.
% It shows that both accesses the less recently tree.
% The intersection of such entry is the first eviction candidate after prime}
%
Equivalently,
the attacker then is always retouching the
PLRU cache way,
no matter which side the sub-tree resides on
in the Tree-PLRU structure.

An example where the attacker applies PLRU Aware \retouchonly
to the case in ~\autoref{f:false-positive} is shown in~\autoref{f:equalizing}.
As we can see,
the attacker chooses to retouch \WC{0} 
instead of \WC{1} as in~\autoref{f:false-positive}. 
When the synchronized sequence \WC{A}
takes place on the left side,
\WC{0} is first evicted by the victim prefetch (\BC{2})
as the resulting PLRU cache way immediately after attacker prime (\BC{1}),
and then brought back to the PLRU cache way by the attacker retouch as a cache miss (\BC{3}).
On the other hand,
when unsynchronized sequence \WC{E}
takes place on the right side,
\WC{0} is accessed by the attacker retouch (\BC{2})
as the resulting PLRU cache way immediately after attacker prime (\BC{1}).
In both cases,
the attacker successfully
retouches the PLRU cache way.
As a result,
sequences \WC{A} and \WC{E}
are nicely distinguished by
examining the PLRU cache way in the
final tree states.

\PN{Reduction to synchronized \sys.}
% only when 
% retouching happens after prefetch 
% so that eviction tree is manipulated as wish in synchronized case.
The insight brought by PLRU Aware \retouchonly
is that prefetch is essentially a retouch with
an eviction of the PLRU cache way
in the perspective of the Tree-PLRU structure.
Essentially,
PLRU Aware \retouchonly,
as demonstrated in~\autoref{f:equalizing},  
reduces unsynchronized operation sequence \WC{E}
to synchronized sequence \WC{B}. 
Similarly,
unsynchronized operation sequence \WC{D}
is reduced to unsynchronized operation sequence \WC{A}.
Consequently,
the final tree states can be analyzed in the same way as
the synchronized case~\autoref{ss:leaky-tplru}.
Furthermore,
the reduction from unsynchronized \sys
to synchronized \sys 
offloads the synchronization analysis to
a post-transaction phase
and therefore eliminates any runtime checking
of victim prefetch that can introduce extra noise
to the attacker. 

\begin{figure*}[th!]
\centerline{\includegraphics[width=0.95\textwidth]{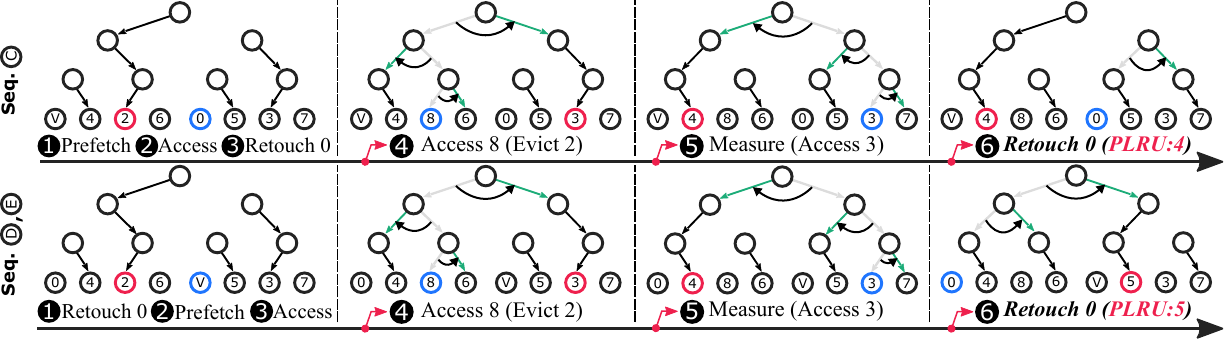}}
\caption
{
	Although the same PLRU cache ways are produced
	after applying PLRU Aware \retouchonly,
	victim entries \protect\WC{V} in 
	sequence \protect\WC{C} and sequence \protect\WC{D}
	end up in opposite sub-trees in the Tree-PLRU metadata.
	By retouching an attacker's entry
	in the opposite group, \protect\eg, \protect\WC{0}, 
	after probing the PLRU cache way (\protect\WC{2}),
	different new PLRU cache ways are produced.
	Consequently,
	the two sequences are differentiated
	by probing the new PLRU cache ways again (\protect\WC{4} \cc{vs.} \protect\WC{5}).
}
\label{f:shifting}
\end{figure*}

\subsection{A poor attacker's approach with delay reduction}
\label{ss:poor-man}
\PP{Challenge: if retouching occurs too late, 
attacker can still miss the victim's access to the prefetched data}
This is because the PLRU cache way resulting from 
sequences \WC{B} and \WC{C} are the same
as shown in~\autoref{t:five-case}, 
so that the attacker can miss a genuine victim access 
if the sequence \WC{C} has occurred, 
producing \emph{false negatives}. 
Note that the result is not dependent on retouching method.
Furthermore,
this challenge can be imposed 
even when retouching is synchronized after the victim's prefetch.
Even the synchronization support such as \primeabort
only guarantee that the retouching happens after the victim's prefetching
not before the victim access if presents. 
Further analysis is thus required to 
capture missed accesses to avoid false negatives. 

\PP{A Poor man's algorithm}
The intuition of the poor man's algorithm is
to locate the attacker's retouching as closely as possible
to the victim's prefetching so that
the victim's access is unlikely to occur in between.
The attacker first sets a long delay after the initial prime
so that the attacker's prefetching
is guaranteed to occur after the victim's access
if there has been one.
Then,
the attacker keeps
relaunching the \sys attack while
reducing the delay in a fine-grained manner
until the unique PLRU entry of sequence \WC{A} is captured,
which confirms a victim access,
or the delay reaches zero,
indicating sequence \WC{B} has occurred instead.
To successfully achieve this, two prerequisites are required.
First, the poor man's retouching should always be synchronized 
after the victim's prefetch. Second, the delay reduction 
should be fine-grained enough to
make retouching intervene the prefetch and possible genuine access
so that it can reveal the victim's access hidden by the false negative.

\PP{Synchronization recovery for poor man's algorithm}
Although PLRU Aware \retouchonly eliminates false positives, 
it does not guarantee that the attacker's retouch always occurs after the victim's prefetch,
a prerequisite for poor man's algorithm in~\autoref{ss:poor-man}. 
However,
as the attacker does not have any timing information
or convenient tools such as Intel TSX,
extra analysis is needed to tell the order of occurrence.
After the desired order of occurrence can be always achieved, 
we can apply the poor man's algorithm 
to further eliminate false negatives.
In short,
we need to differentiate sequence \WC{C} from \WC{D}
(we omit sequence \WC{E}
as its behavior replicates \WC{D} due to
back-to-back access to \WC{V}):
\begin{flalign*} 
&\WC{C} Prefetch_{Victim} \rightarrow Access_{Victim} \rightarrow Retouch_{Attacker}  \\
&\WC{D} Retouch_{Attacker} \rightarrow Prefetch_{Victim} \rightarrow Access_{Victim} \\
&(\WC{E} Retouch_{Attacker} \rightarrow Prefetch_{Victim})
\end{flalign*}
More specifically,
if we can distinguish the two sequences
we can recover the synchronization information about 
the victim's prefetch and the attacker's retouch.
Although the final state of the tree remains
after applying PLRU Aware \retouchonly,
it is worth noting that
depending on the time when retouching occurs,
the victim cache line ends up in the opposite group of the Tree-PLRU structure,
\autoref{f:shifting},
after probing the PLRU cache way (\BC{5}).
Namely,
the root flag directs towards the victim entry under sequence \WC{C}
and leads away from the victim entry under sequence \WC{D}.
The same situation applies to the Eviction Aware \retouchonly-ed cache line symmetrically.
This fun fact can later be utilized by the attacker
to recover the occurrence order of
the victim's prefetch and the attacker's retouch,
avoiding the noise brought by probing
the eviction metadata during runtime for such information.

As shown in~\autoref{f:shifting},
after probing the PLRU cache way (\BC{4} and \BC{5})
following the outcomes in ~\autoref{f:equalizing},
victim entry \WC{V} resides in opposite
Tree-PLRU groups (\BC{5}) with the same new PLRU cache way (\WC{4}).
By retouching the previously Eviction Aware \retouchonly-ed cache line, \WC{0} in this case,
sequence \WC{C} will not cause the next PLRU cache way (\cc{PLRU:4}) to change,
as the root flag leads away from \WC{0},
while sequence \WC{D} alters the next PLRU cache way (\cc{PLRU:5}) as the root flag directs towards \WC{0}.
We then can differentiate the two sequences 
by probing the eviction candidate again,
and recover the occurrence order.

% WE NEED AN APPROACH THAT CAN SOLVE BOTH AT THE SAME TIME. 
% Although we can solve first problem with deferring measurement to later we cannot ensure that retouching has happend in between prefetch and real access. 
% However, in our modified scheme, it evicts 3 only when prefetch is located in between two accesses
% Therefore, by reducing delays, first we can find a point that next victim will be changed a lot
% And then by probing those timing region in some threshold times, we can more precisely verify whether there has been actual access or not.

%need to measure that re-access happens after victim's prefetch
%need also meausre that re-access happens before victim's access

%these two measures are required at every execution because memory accesses latency inside the victim process can vary 
%even with the same cache structures and same cache lines are located in the same hiearchies. 
%there is no guarantee that the same latency will remain in the next run

\section{\sys on X86}
\label{s:eval}
\PP{Responsibility disclosure}
We have reported \sys to Intel
and received an acknowledgment.

\PP{Execution environment}
Our evaluation platform is equipped with
an Intel Core i9-9900K (Coffee Lake) CPU @ 3.60 GHz,
running Linux Ubuntu 20.04.1 LTS with a 5.4.0-rc8 kernel.
The size of the L1 data cache is 32KiB per core, and
it consists of 64 cache sets with
eight cache lines per set (8-Way).
The size of cache line is 64Bytes.
The machine uses microcode version 0xd6 and adopts \emph{R0 stepping (\ie, 13)}.
Therefore, we assume that defense described in \autoref{ss:limited-sync} is available, and
the attacker cannot utilize TSX to assist synchronization.
We assume that the hyperthreading is enabled and
the attacker can co-locate a spy process concurrently with the victim process
so that the L1 cache is shared.
Last, we assume that attacker can utilize \emph{rdtsc} 
to capture the L1 miss event. 

\PP{Prefetching and locking defense mechanisms deployed}
For data preloading to L1 cache, we utilize \emph{write} operations
inside a TSX transaction.
Note that \emph{write} operation is required because of 
implementation of~\cite{cloak}, not because of intentionally launching the \sys
on a weak assumption.
The victim accesses preloaded entries
as he wishes using \emph{read} operations.
We assume that 
security-critical code and data
can fit and are properly placed inside 
the TSX protection regions.

\PN{\sys in SGX}
For SGX environment,
we exclude specific cores from scheduling
by applying the \cc{isolcpus with noHZ} boot parameter of the Linux kernel
and use \cc{rdpmc} instruction to
capture L1 cache miss events
to reduce noise.
If there is no mention about SGX explicitly, 
all the experiment results are retrieved 
in a non-SGX environment.

\subsection{Baseline Evaluation of \sys}
To effectively demonstrate the basic capability of \sys,
we implemented a simple version of~\cite{cloak}.
The victim process prefetches 64 data blocks,
each mapped to a 64 different cache set 
inside a TSX transaction
(L1 data cache consists of 64 sets in x86).
We locate the victim's access operation
after the prefetch based on the experiment settings.

\begin{figure}[!t]
\centerline{\includegraphics[width=\columnwidth]{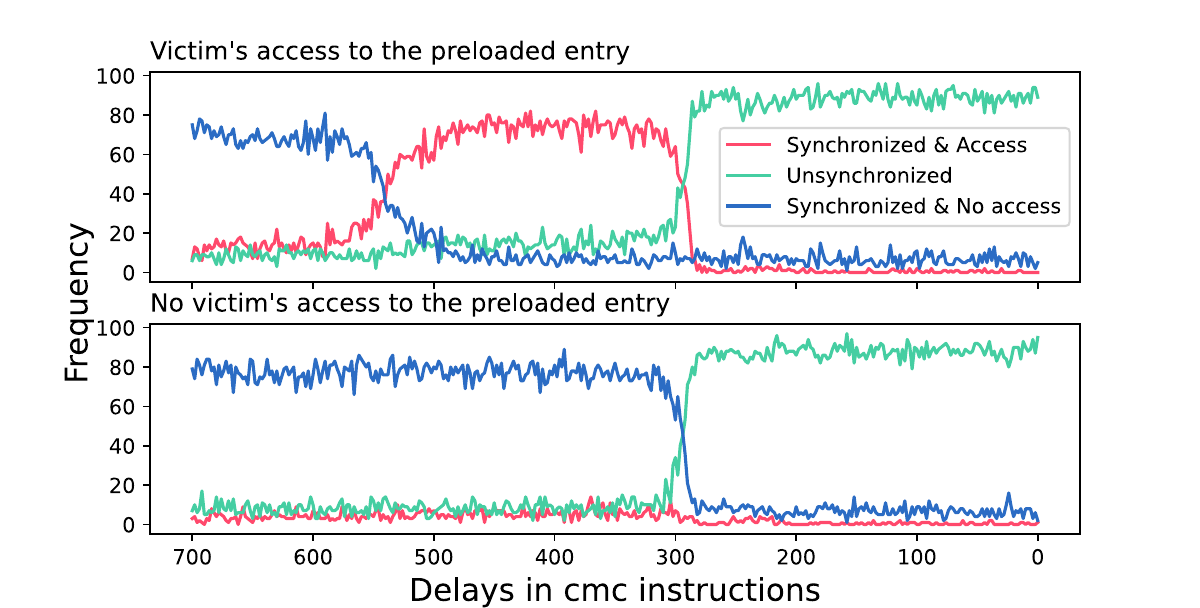}}
\caption{
	The only difference in these two graphs 
	is the presence of victim's access after the prefetch. 
	}
\label{f:poor-man}
\end{figure}

\PN{PLRU Aware \retouchonly can effectively distinguish all operation sequences}
We demonstrate that 
\sys can accurately distinguish victim's accesses
by leveraging PLRU Aware \retouchonly in~\autoref{f:poor-man}. 
In addition, 
we reveal the synchronization information 
retrieved from the Tree PLRU 
that allows the attacker to estimate 
when a victim's further access occurs if it exists.
In this experiment,
we launched \sys 100 times per delay
from 700 to 0 \cc{cmc} range with \emph{2 \cc{cmc} instructions} 
as cycle reduction granularity.
Also, we iterated the identical experiment for two cases:
when the victim's access follows the preloading (above graph) 
and when no access follows the preloading (below graph).
For each experiment, 
\sys successfully distinguished all three cases: 
(1) retouching in between prefetch and access (\emph{Synchronized, access}),
(2) retouching before prefetch (\emph{Unsynchronized}), and
(3) retouching after prefetch (\emph{Synchronized, no access}).
As shown in the upper graph, there is a distictive increases 
on the \emph{Synchronized, access} case around 550 to 330 \cc{cmc} delays.
This indicates that victim's genuine access occurs 
in that interval most of the time. 
However, the lower graph couldn't capture any drastic frequency changes 
in the same case (Red), which means no victim access was initiated after the preloadindg.
Also, an intersection of two different lines at around 300 \cc{cmc} in both graph 
demonstrates that the time of victim's prefetch operation can also be pinpointed 
with \sys.

\begin{figure}[t]
\centering
\begin{tikzpicture}
\begin{axis}
    [
        xmin = 00,
        xmax = 700,
        ymin = 0,
        ymax = 100,
        xlabel = Number of \cc{cmc} instructions,
        ylabel = \shortstack{Number of\\monitored access},
        width=\axisdefaultwidth,
        height=\axisdefaultheight*0.4,
        x dir=reverse,
        legend style={nodes={scale=0.5, transform shape}}
    ]

    \addplot[line width=0.7pt,color=lcolor1,smooth] %
      table[x=x,y=y,col sep=space]{data/back-to-back/10.dat};
    \addplot[line width=0.7pt,color=lcolor3,smooth] %
      table[x=x,y=y,col sep=space]{data/back-to-back/30.dat};
    \addplot[line width=0.7pt,color=lcolor4,smooth] %
      table[x=x,y=y,col sep=space]{data/back-to-back/50.dat};
    \addplot[line width=0.7pt,color=lcolor5,smooth] %
      table[x=x,y=y,col sep=space]{data/back-to-back/60.dat};
    \addplot[line width=0.7pt,color=lcolor6,smooth] %
      table[x=x,y=y,col sep=space]{data/back-to-back/63.dat};
    \legend{Set 10, Set 30, Set 50, Set 60, Set 63}

\end{axis}
\end{tikzpicture}
\caption
{
    Reliable false negative reduction achieved by poor man's algorithm.
    \emph{1 cmc instruction} is set as the granularity of delay reduction.
    The initial delay has been set as \cc{700 cmc} instructions, and
    we iterated 100 times per delay until it reaches \cc{0 cmc} delay.
}
\label{p:back-to-back}
\end{figure}
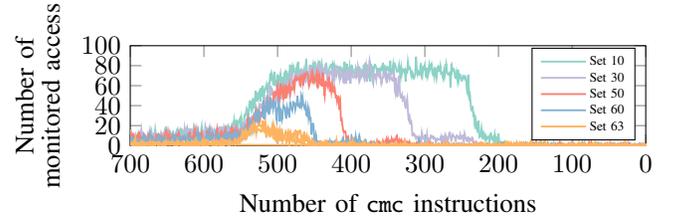

\PN{Poor man's algorithm reliably make \retouchonly to be located between victim's prefetch and access} 
When the victim's access occurs after preloading,
timing delay between
the preloading and actual access
can determine the accuracy of \sys.
It is easier to capture the access
when the victim touches the entry 
prefetched earlier
because there is sufficient timing delay 
that attacker can intervene for \retouchonly.
For address mapped to cache set N ($0 \leq N \leq 63$),
it has around $(63-N) * prefetch\_latency$ delay
between the preloading and the further access.
%
%Note that this characteristic of \sys is presented in~\autoref{f:retouch64}
%(detection rate drops to 20\% for the last set).
%
To further evaluate the reliability of \sys,
we launched \sys against five different cache sets
that have various timing intervals between the prefetch and the following access.
Our result shown in~\autoref{p:back-to-back}
proves that the poor man's algorithm described in~\autoref{ss:poor-man}
reliably eliminates false negatives and accurately identifies the victim access.
Although the signal telling presence of genuine accesses is getting weaker, 
as the time interval between the prefetch and the access decreases,
it is enough to distinguish the victim accesses from the noise.

\subsection{Attacking AES T-Table}
\label{ss:eval-impact}
%T-Table is already known to be vulnerable but exists for cache side channel comparisions in lots of researches 
Although appropriate countermeasures~\cite{gueron2010intel, kasper2009faster, rebeiro2006bitslice,konighofer2008fast}
have closed most of the cache side-channels presented in AES implementations,
still the T-Table implementation of AES is frequently used 
to explore different characteristics of new side-channel attacks
compared to the existing approaches ~\cite{saattack, gruss2016flush+, gruss2015cache,cloak, reload+}.
Therefore, 
we demonstrate the effectiveness and stealthiness of \sys compared to \primeprobe
by attacking T-Table based AES encryption,
both in regular environments and in \SGX enclaves.
The T-Table implementation transforms 
AES operations (\ie, SubBytes, ShiftRows, and MixColumns) in one round of encryption
into 16 memory lookups to 4 different T-Tables.
The T-Table accesses in the first round of encryption are stated by the equation 
$T_{j}[p_{i} \oplus k_{i}]$ with $i \equiv j \mod 4$ and $0 \leq i < 16$.
Therefore,
if the attacker can infer the values of
$p_{i} \oplus k_{i}$, the indexes to T-Tables,
they can then narrow down the possible key-bytes ($k_{i}$)
in case the plaintext ($p_{i}$) is known~\cite{prime-probe-dcache, advances, bernstein2005cache}.

\begin{figure}[!t]
\centerline{\includegraphics[width=\columnwidth]{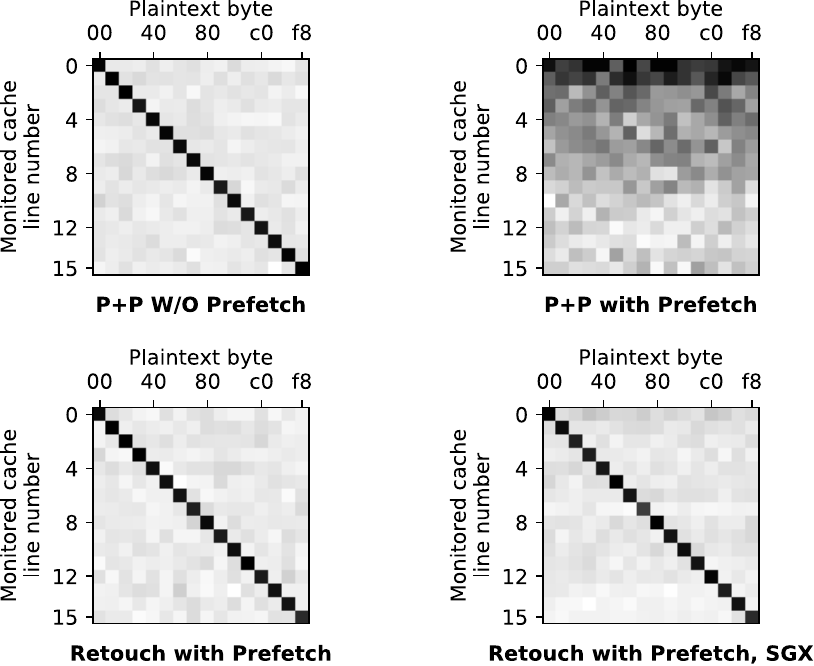}}
\caption{
    Color matrix showing cache hits on AES T-Table encryption. Darker means more L1 cache hits. 
    P+P indiciates \primeprobe. 
    We conducted four different experiments on the same AES implementation. 
    The first implementation was conducted with \primeprobe without the prefetch-based defense applied.
    The other experiments were performed with both \primeprobe and \sys,
    with prefetch-based mitigations applied.
}
\label{f:AES}
\end{figure}

\PP{AES settings}
Our experiment assumes the known-plaintext attack scenario
and targets the AES implementation in OpenSSL.
Each T-Table consists of 256 entries of 4bytes (1KB of each) 
and maps to 16 different sequential sets of L1 cache.
The first T-Table, Te0 is always mapped to the page-aligned addresses in our implementation, 
which map to set 0 to 15 of the L1 cache.  
Other T-Tables are mapped to the subsequent L1 data cache sets (16 to 63).
We arranged the plaintext bytes 
to make the AES encryption access the cache line 
equal to $p_{0}$ {\small/} $0x10$. 
The plaintext value shown in the ~\autoref{f:AES} 
indicates the first byte of plaintext ($p_{0}$), and 
the random values are assigned to the left 15 bytes of plaintext ($p_{1-15}$). 
Also, we generate a 128-bit AES key with zero-filled user key data.
Note that this experiment setting is general
and adopted in previous researches ~\cite{gruss2016flush+, cloak, gruss2015cache}.

\PP{Evaluation methods}
We launched around 300,000 encryptions asynchronously
and measured the victim's access 
made on the first T-Table mapped to the L1 cache set 0 to 15.
For the prefetch-based defense, 
we first preloaded the entire four T-Tables to the \TSX \cc{write set}
and performed ten rounds of AES T-Table encryptions 
in a single TSX transaction. 
To bring the T-Table entries into a \cc{write set}, 
we changed the T-Tables to be writable and introduced a minimum of 64 write operations, 
each of which strides cache line size.
We measured the average time 
to preload 64 memory entries with \cc{write} operations 
and organized the same amount of delay composed of \cc{cmc} instructions.
As a result,
we could make the \retouchonly operation intervene 
between the preloading and actual AES encryptions. 
Also, we put additional random delays of less than 300 \cc{cmc}
to make the measurement operations initiate 
after the first round of AES encryption. 
The identical \sys attack was also conducted in \SGX enclaves
with the prefetch-based defense applied.

\PP{Evaluation analysis}
As shown in~\autoref{f:AES}, 
without prefetch-based mitigations, 
\primeprobe attack is sufficient 
to precisely leak the entire T-Table access patterns.  
However, when preload operations load the entire T-Table to the L1 cache sets 
before the actual AES encryption, 
\primeprobe can no longer distinguish real access patterns. 
The leakage captured by \primeprobe under the prefetch-based defense shows 
as if the victim process accessed entire T-Tables 
during the AES encryption operation. 
%%%%%
However, 
we can accurately distinguish the actual T-Table accesses with \sys
even when prefetch-based defenses are applied,
both in regular environments and in \SGX enclaves.
Note that the \primeprobe without prefetch and \sys with prefetch
have a similar result, proving that \sys is as effective as traditional
cache side-channel attacks in regular environments as well as in \SGX enclaves.
In both attacks, we can clearly see the diagonal cache access traces.
Also, 40\% to 60\% of the victim's accesses are captured for the other cache lines 
because we randomize the $p_{1-15}$.
This indicates that \sys is accurate enough
to capture actual access patterns 
even under the prefetch-based mitigation, 
which is not possible with eviction-based side-channel attacks.
%%%%
Another advantage of \sys is stealthiness.
we observe that only 2.169\% of
the executions incur TSX faults while launching \sys. 
Among all TSX faults, 
10\% were due to cache misses (\ie, 0.2\% of all executions)
while the rest were caused by other TSX faulting conditions,
leaving minimal attack traces and making the attack hardly detectable.
This indicates that most of the TSX faults were caused by 
running AES encryption in one transaction and irrelevant to the attack with \sys.
To verify it, 
we launched the same AES implementation without involving the \sys attack, and 
observed again that around 2\% of the executions incurred TSX faults.
In contrast, 
when we launched the traditional \primeprobe attack 
against the same AES implementation 
protected by the prefetch-based defense, 
93.445\% of the executions triggered TSX faults.

\section{\sys against M1}
\label{s:eval-m1}
\PP{Execution environment}
We used the Apple Mac Mini with the M1 CPU
consisting of four performance cores (Firestorm) and
four efficiency cores (Icestorm).
Each Firestorm core has 128KB of L1 data cache;
and each Icestorm core has 64KB of L1 data cache.
The machine is equipped with 16GiB of memory and runs
a custom build Ubuntu 5.11.0-rc4+.
We assume that the victim adopts naive prefetching without locking because
M1 has no support for transactional memory.
However, we assume that the victim's entry should not be evicted 
while launching the \sys.
Also, we assume that the victim and attacker processes run
under time-sliced sharing setting such as pthread or user-level threading.
Last, we assume that the attacker has no root privilege.

\subsection{Challenge of \sys on M1}
To determine whether the victim has accessed the prefetched data,
the attacker should be able to utilize PMU or high resolution timer
as L1 cache activity monitor. 
However, 
unprivileged processes (running at EL0) are not authorized 
to use those measurement tools by default on the M1 platform. 
%
%Therefore, the attacker requires another measurement tool
%that can capture the L1 cache miss event.

\PP{Interface of ARM system counter is hidden}
ARMv8 architecture employs a system counter~\cite{arm-generic-timer}
which provides a fixed frequency incrementing system count 
giving an equivalent view of the passage of time.
Fortunately, the system timer is accessible to unprivileged users
through the ARM-recommended register interface, \cc{CNTVCT_EL0}.   
However, we found that reading the system counter using this interface raises 
an illegal instruction fault on the M1 platform.
Nonetheless, we could discover that an undocumented system register 
\cc{s3_4_c15_c10_6} is used as an alias of \cc{CNTVCT_EL0}.
To locate this unknown interface,
we cleverly leveraged Apple's Rosetta2~\cite{rosetta2} which translates 
x86 instructions to the ARM64 counterparts.
Because x86 provides a user-accessible system timer 
through \cc{rdtsc} instruction,
we made Rosetta2 translate the x86 binary containing \cc{rdtsc} instruction 
and disassembled the translated ARM64 binary.
As a consequence, we can ascertain that the \cc{s3_4_c15_c10_6} is utilized 
to access system counter at EL0.

\begin{figure}[!h]
\centerline{\includegraphics[width=\columnwidth]{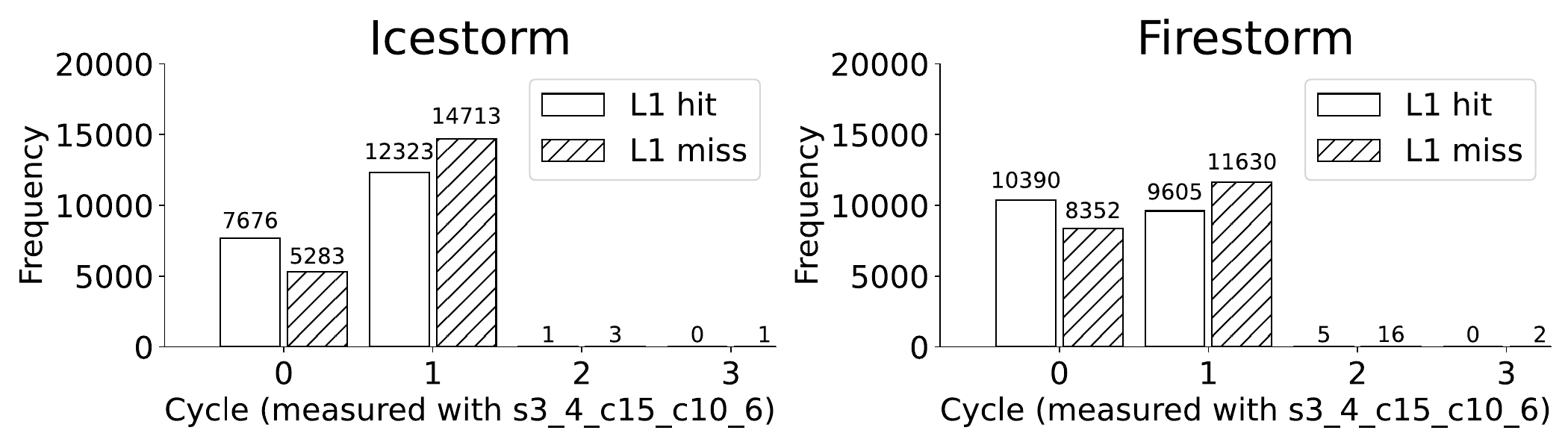}}
\caption{
	Latency of a \cc{ldr} measured with the coarse-grained system timer. 
	The measurement iterated 20000 times for L1 hit and L1 miss respectively.
	0 cycle indicates that the latency of load operation is too small to be measured with the system counter.
        }
\label{f:m1-counter-miss1}
\end{figure}

\begin{figure*}[t!]
\centerline{\includegraphics[width=1.00\textwidth]{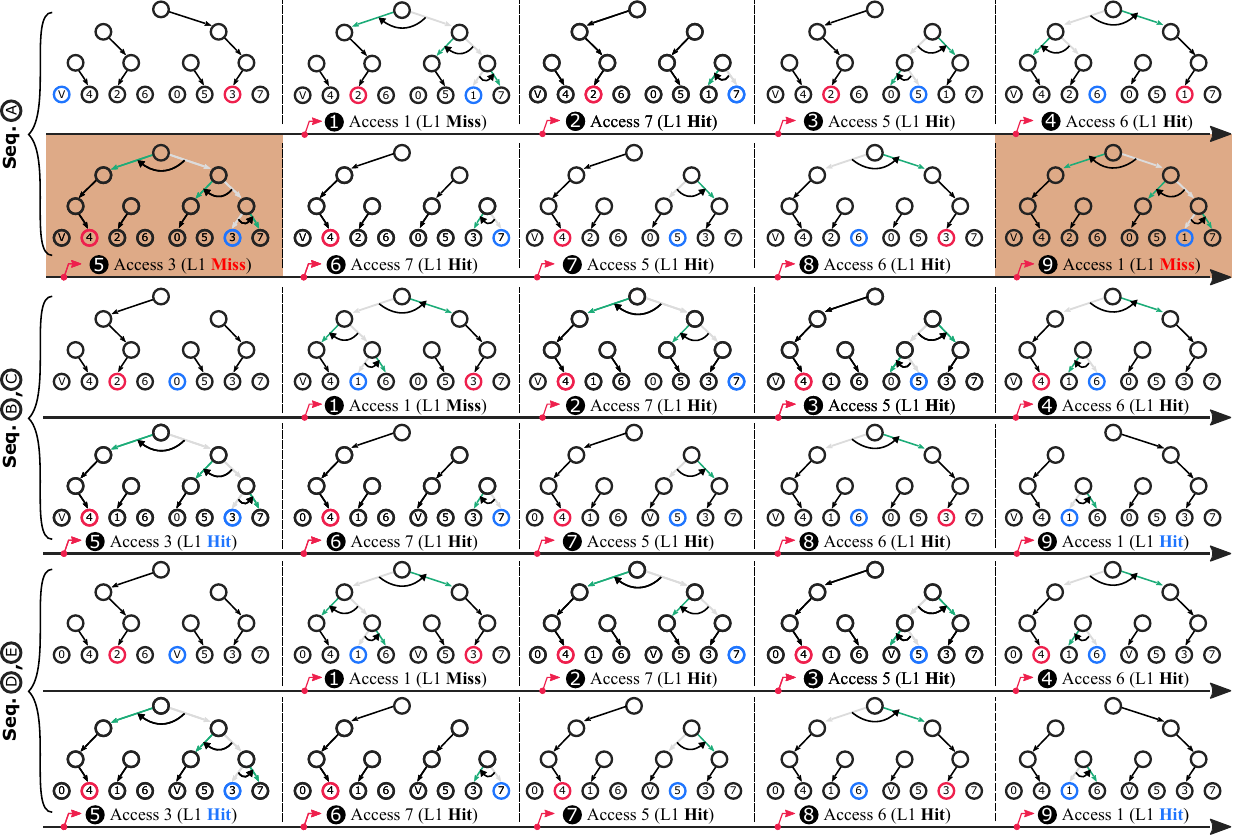}}
\caption
{
	Access sequence designed for introducing L1 miss events only for the operation sequence \protect\WC{A}.
}
\label{f:shifting-250mhz}
\end{figure*}

\PP{System counter is too coarse-grained to differentiate between L1 hit and miss events}
In~\autoref{s:m1-reversing},
we confirmed that M1 architecture exhibits a clear distinction between
the L1 hit and miss latency.
The latency differences are around 10 and 12 core clock cycles 
on Icestorm and Firestorm, respectively~\autoref{f:m1-l1-latency}.
However, 
it is known that Firestorm core runs at 3.2GHz, but 
the system counter increases with a frequency in the range of 1MHz to 50MHz
according to~\cite{arm-generic-timer}.
Therefore, the two events cannot be reliably distinguished 
due to the coarse granularity of the system timer.
To validate the claim, we measured the elapsed time for accessing 
one memory address using the system timer.
%We used the same load sequences that were utilized in~\autoref{f:m1-l1-latency}
%to make the memory operations served by the L1 cache; 
%and the others from the L2 or L3 cache.
Because we have full knowledge about the eviction policy of L1 data cache,
we can easily produce L1 hit and miss conditions on the measurement. 
As shown in~\autoref{f:m1-counter-miss1}, 
there is no clear distinction in between the two events. 
When load operation finishes at 1 system clock cycle, 
it can be either L1 hit event or L1 miss event with 45.57\% and 54.42\% chance, respectively.
For Firestorm, it was 45.23\% and 54.76\%.
Note that the similar ambiguity is also present in the 0 cycle case in both cores.
In summary,
the latency difference between the L1 hit and miss event is too negligible 
to be distinguished with the coarse-grained system counter.
Therefore, we need another approach to infer victim's access in \sys under M1. 

\subsection{Amplifying weak signal with Tree-PLRU}
\label{s:m1-overcome}
We revise \sys by leveraging characteristics of Tree-PLRU
so that the attacker can deliberately produce more miss events
only when a particular operation sequence happens.
Although the latency of a single miss event cannot be captured,
when more miss events are accumulated, it will become large enough to be
measured with the coarse-grained system timer.
%
%\PP{Amplifying the signal with more misses}
To infer which operation sequence happens,
the attacker has to make one operation sequence always produce
more L1 cache miss events than the others in the measurement phase.
In detail,
the attacker issues a particular memory access sequence designed 
to introduce additional L1 miss events only for the operation sequence \WC{A}
where the victim accesses the prefetched data.
%More miss events can accumulate the latency difference between a miss and hit event of that sequence, 
%resulting in noticeable latency differences that can be measured by the system timer. 
%
It is worth noting that 
if the access sequence introduces the same amount of L1 misses 
to all other operation sequences (\WC{B} to \WC{E}),
the attacker cannot tell which operation sequence has occurred.

\PP{Exploiting Tree PLRU state to produce more L1 cache misses}
It is non-trivial to generate an access sequence
satisfying such a condition
because the attacker should consider additional state changes
made on the other result sequence together.
The major principal in constructing the proper memory sequence is 
maintaining at least one different entry in all possible Tree-PLRU states
after every access.
When we have a different entry, 
we can easily generate different cache events
for the same memory accesses 
because one operation sequence has entries 
that are not owned by the others and vice versa. 
However, as shown in~\autoref{f:shifting-250mhz}, 
initially, all five operation sequences have the same cache entries. 
Nonetheless, we can satisfy the principal 
because the Tree-PLRU state of sequence \WC{A} and the others 
designate different entries as the next eviction target
due to the PLRU aware \retouchonly.
In step \BC{1}, accessing cache line \WC{1} makes any operation sequence to produce miss events.
However, it makes only the sequence \WC{A} own different entry \WC{2} compared to the others
(note that \WC{2} is evicted from the others and \WC{3} remains instead).
After that, through steps \BC{2}-\BC{4}, 
we should manipulate the Tree-PLRU state of \WC{A} so that 
the further L1 miss events always replace the entry recently brought to the set (\WC{1}).
Under such Tree-PLRU settings, fetching \WC{3} will result in a cache miss to \WC{A},
but will result in a cache hit to the others (\WC{B}-\WC{E}).

\begin{figure}[!t]
\centerline{\includegraphics[width=\columnwidth]{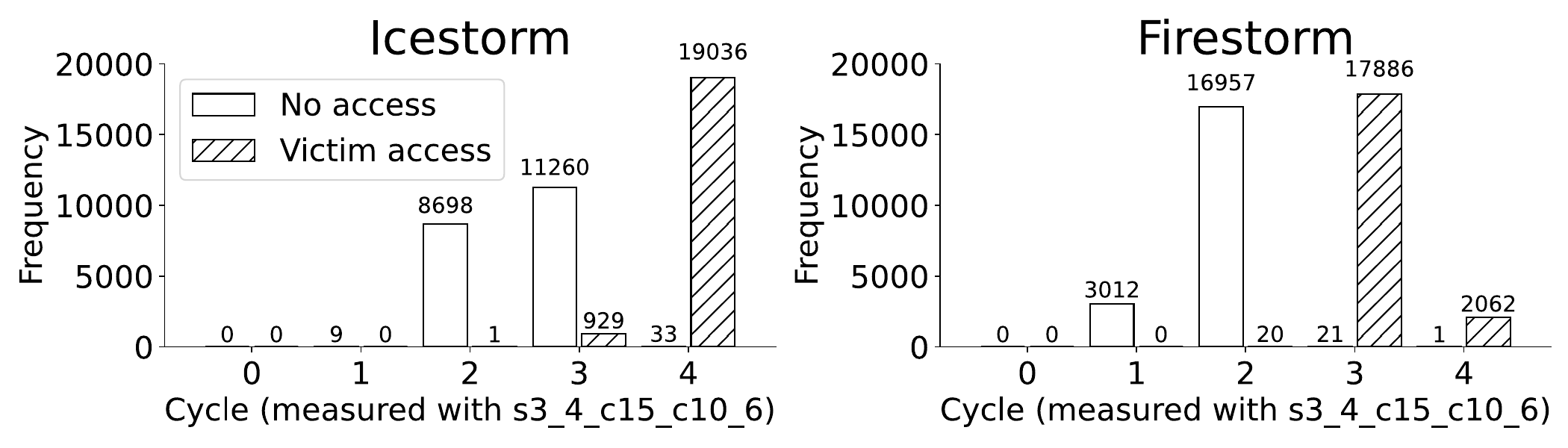}}
\caption{
	The access sequence comprises of 57 load operations 
	repeatedly accessing cache entries in the following order $\overline{75637561}$.
	It makes the sequence \protect\WC{A} (case of victim access)
	produce 14 additional misses compared to others.
        }
\label{f:m1-counter-miss14}
\end{figure}
\PP{Amplified L1 miss event can distinguish victim's access} 
To show that the system timer can sufficiently capture the
victim access with the L1 miss signal amplification technique,
we implement a simplified victim program that prefetches 
single data block mapped to a particular L1 data cache set.
Also, to minimize the noise induced by time-slicing and 
show the effectiveness of \sys in M1, 
we utilize synchronization primitives to yield the victim program
for the attacker to retouch in time. 
The attacker launches \sys and 
produces additional memory access sequences
described in \autoref{f:shifting-250mhz} in the measurement.
We empirically found that 14 additional misses are enough to clearly distinguish
genuine victim accesses in both types of core. 
We can produce one additional miss every four memory access operations.
Therefore, to introduce 14 additional misses, 
we added 57 additional load operations including the first access(\BC{1}).
Also, we measure the latency to execute 57 additional load misses with the system timer.
Note that the miss happens only when the sequence \WC{A} happens.
Therefore, as shown in \autoref{f:m1-counter-miss14},
only when the victim accesses data after the prefetch (hatched bar plot)
it incurs additional clock cycles to run 57 load operations.
Compared to result shown in \autoref{f:m1-counter-miss1} that measure 
only one miss event with the system timer,
our results clearly distinguish the genuine victim access. 
Although we demonstrated \sys on the restricted environment
to show its effectiveness,
we believe that the attack can be reproducible 
against realistic examples.

\section{Related work}
\label{s:relwk}
\PP{Tree-PLRU cache eviction policy abuses}
\cite{l1covert} establishes covert channel 
by encoding one-bit information with the next eviction target
based on the Tree-PLRU eviction policy.
Also, ~\cite{DAWG} theoretically shows how the shared Tree-PLRU state
can break the security of cache partitioning defenses.
A noticeable difference between \sys and ~\cite{l1covert, DAWG} is that
it can precisely recognize actual accesses even under the presence of preloading and locking defense.
In detail, \retouchonly operation on an attacker's primed cache line manipulates the
Tree-PLRU state in such a way that a following victim's access reveals its existence. 
However, under the same condition, 
previous attacks cannot capture the victim's access
because prefetch-based defense eliminates the leakage. 
Furthermore, 
the previous attacks have not been fully demonstrated with 
realistic and detailed examples regarding the Tree-PLRU eviction policy.
Also, \sys provides additional synchronization information 
for the attacker to reduce attack noises, 
which has not been explored in the previous works. 

\PN{Quad-LRU cache eviction policy abuses}
\cite{reload+} shows that the Quad-age LRU (QLRU) eviction policy~\cite{qlru}
can be exploited to leak the victim's access pattern from the L3 cache.
In essence, 
the QLRU records the number of accesses 
made to a specific entry with a 2-bit counter
dedicated to each entry. 
Due to this property of QLRU, 
consecutive accesses to one cache line back-to-back 
can be clearly observed as two separate accesses. 
However, the Tree-PLRU can track only the access order among the cache entries,
not the number of accesses.  
Therefore, it is more challenging to distinguish the victim's access 
when the entries are prefetched before the access. 
\sys shows that \retouchonly allows the attacker to capture the victim's access 
despite the restriction of Tree-PLRU eviction policy.
Furthermore, \sys does not evict the victim's cache line, 
but \cite{reload+} can induce such evictions 
depending on the victim's accesses.
When preloading and locking defense is deployed, 
eviction of the victim's cache line will be detected immediately, 
which can make the reloading operations meaningless.
Note that this difference makes the \sys more stealthy.
Lastly,\cite{reload+} requires shared memory, which is not a requirement of \sys.

\PP{Apple M1 on Asahi Linux}
\cite{asahi-linux} is an ongoing project that aims at running Linux operating system on Apple M1 architecture. 
As part of the porting, they also provide a list of Apple's system registers, 
including configuration registers for Apple's proprietary PMU.
However, their Linux version does not include PMU support at the time of submission. 
Thus, to the best of our knowledge, this is the first work that fully demonstrated 
how the Apple PMUs and its undocumented events can be utilized to reveal the underlying cache architecture.

\PP{Google's Tree-PLRU attack}
As a concurrent work, 
Google showed that a single read of secret data is enough 
to leak data efficiently by abusing the characteristic of the Tree-PLRU cache eviction strategy 
in their blog post~\cite{google-similar}.
Compared to Google's work, 
\sys further demonstrates that Tree-PLRU cache eviction side channel 
can even efficiently bypass the prefetching-and-locking defenses.
Also, we provide in-depth micro-architecture analysis on Intel x86 and Apple M1 architectures, 
lacking in Google's article, which validates our claims.
More importantly, we reported our attack to Intel and received an acknowledgment about the attack method 
before the Google's publication\footnote{We first reported our attack to Intel on Nov 3, 2020, and 
Google posted the article on March 12, 2021.}.

%possible mitigation

%fill the cache set with the victim's entry fully. 
\section{Conclusion}
One of the most important characteristics of \sys 
that differentiates it from other cache side-channel attacks is 
full awareness of eviction policy, especially about Tree-PLRU.
Based on the comprehensive understanding of Tree-PLRU policy
we designed PLRU Aware \retouchonly
that can completely break prefetch and locking defense.
We demonstrated our attack is feasible 
by showing the leakage in AES T-Table encryption, 
which cannot be achievable in the 
eviction based side-channels.
Furthermore, 
we demonstrates that \sys can leak victim's access pattern
even without access to the fine-grained timer
on the M1 platform.
Also, we first provide detailed analysis of L1 data cache 
on M1 platform.

\clearpage
%\footnotesize
%\setlength{\bibsep}{3pt}
\bibliographystyle{IEEEtranS}
\bibliography{p,sslab,conf,cache,sgx}

\clearpage
%\nobalance
\appendix
\renewcommand\thefigure{\thesection.\arabic{figure}}
\setcounter{figure}{0}
\section{Appendix}
\label{s:appendix}

\subsection{Implementation of reversing L1 cache on M1}
\label{ss:m1-reverse}
\begin{figure}[!htb]
 \input{code/m1.c}
	\caption{Basic reversing code for L1 DCache on M1. 
	\cc{PMC2} and \cc{PMC3} has been set to track core clock cycle counter and L1 data cache miss event, respectively}
 \label{f:code-m1}
\end{figure}
\PP{System register configurations for Apple's PMU}
To use the Apple's PMU, several system registers should be correctly configured.
The \cc{PMCR0} register mapped to \cc{s3_1_c15_c0_0} controls basic capabilities of PMU such as 
which Performance Monitoring Counter (PMC) needs to be enabled (bit 0-7), which interrupt mode of PMUs will be used (bit 8-10), and 
authorizing unprivileged user's access on the PMU (bit 30). 
By default, the bit dedicated to enabling user access on PMU is disabled on macOS 
so we also disabled that bit.
We checked the default \cc{PMCR0} register value on the latest macOS, BigSur 11.4,
by loading the kext module.
Also, because our measurement is done in a very short time window,
there is no need for setting interrupt-related information for PMU control.
Therefore, we just set two bits for enabling \cc{PMC2} and \cc{PMC3} (Line 9).
Apple provides another PMU control register called \cc{PMCR1} mapped to system register \cc{s_1_c15_c1_0}.
This register controls which execution modes count events.
We utilize the Linux driver instead of user process, which runs at EL1 privilege (kernel). 
Therefore, we set the \cc{PMCR1} register to allow our kernel module 
to access the \cc{PMC2} and \cc{PMC3}. 
Because the 8 bits from 16 to 23 is dedicated to controlling
\cc{PMC0} to \cc{PMC7} on the EL1 privilege, we set the two bits as shown in the Line 10.
Lastly, the \cc{PMESR0} register mapped to \cc{S3_1_C15_C5_0} provides a way to 
select PMU events for \cc{PMC2} to \cc{PMC5}.
Each PMC can track PMU events 0 to 254, and 8 bits are assigned per PMC.
The least significant bit 0 to 7 are used to select PMU event of \cc{PMC2}, and 
the next 8 bits are dedicated for \cc{PMC3} event selection.

\PP{Measuring the events using the PMU}
After the PMU registers are properly initialized,
one can measure the event by reading the PMC register 
assigned for tracking specific events.
As shown in Line 39-40 and Line 42-43, 
to measure the events caused by one memory load operation (Line 41),
PMC registers should be read before and after the memory operation
using \cc{mrs} instruction.
Also, to prevent the instruction reordering between \cc{mrs} and \cc{ldr} instruction,
we carefully insert Instruction Synchronization Barriers (ISB).
To reverse engineer the eviction policy of the L1 cache on the M1 platform,
we follow the reversing logic described in~\autoref{alg:rev}.
First, we fill a specific cache set by issuing multiple memory load operations 
to the addresses mapped to the same set (Line 29-33).
And then, we enumerate any combinations of read operations (Line 34-35) 
to change the underlying state of the eviction policy. 
By accessing one more entry mapped to the same set (Line 36-37), 
we can evict one entry from the set. 
After the eviction, we measure the latency of accessing a particular memory address 
we used for filling the L1 cache set (Line 29-33).
Based on the latency difference measured in~\autoref{f:m1-l1-latency},
we can determine which cache line has been evicted and 
recover the eviction policy from the trace.

\end{document}